\documentclass[12pt]{article}
\usepackage[margin=1in]{geometry} 
\usepackage{setspace} 
\usepackage[utf8]{inputenc}
\usepackage[T1]{fontenc}

\usepackage{natbib}
\usepackage{lineno,hyperref}
\usepackage{float}
\usepackage{enumerate}
\usepackage{amsmath}
\usepackage{multirow}
\usepackage{booktabs}
\usepackage{graphicx}
\usepackage{verbatim}
\usepackage[dvipsnames]{xcolor}
\usepackage{subcaption}
\usepackage{amssymb}
\usepackage{pifont}
\usepackage{dirtytalk}
\usepackage{tablefootnote}
\usepackage{amsfonts}
\usepackage{algorithm}
\usepackage{algpseudocode}
\usepackage{accents}
\usepackage{multicol}
\usepackage{soul}
\usepackage{algorithm}
\usepackage{booktabs}

\newcommand{\E}{{\rm E}}
\newcommand{\Prob}{{\rm P}}
\newcommand{\var}{{\rm var}}
\newcommand{\cov}{{\rm cov}}
\newcommand{\corr}{{\rm corr}}

\DeclareMathOperator*{\argmax}{arg\,max}

\newtheorem{theorem}{Theorem}[section]

\newtheorem{corollary}[theorem]{Corollary}

\begin{document}

\begin{center}
{\Large \textbf{A common zero-inflation bivariate Poisson model with comonotonic and counter-monotonic shocks}}

\bigskip

Golshid Aflaki$^a$, Juliana Schulz$^a$, Jean-François Plante$^a$

\begin{footnotesize}
    \noindent $^a$ D\'epartement de sciences de la d\'ecision, HEC Montr\'eal
\end{footnotesize}
\end{center}

\begin{abstract}

There are numerous applications which involve modeling multi-dimensional count data, notably in actuarial science and risk management. When such data exhibit an excess of zeros, common count models are no longer suitable. With multivariate data, characterizing an appropriate dependence structure is equally critical in order to adequately assess the underlying risk inherent in the joint counts. In this work we propose a new bivariate zero-inflated Poisson model appropriate for modeling pairs of counts with a surplus of zeros. The proposed construction is based on a mixture model involving a common mass at zero along with a Poisson random pair. Various forms of dependence are considered for the latent Poisson pair, allowing for both negative and positive dependence. Several model properties are explored, notably the joint probability mass function and implied dependence structure. The method of moments and maximum likelihood approaches are described and implemented for estimation. The usual asymptotic properties of the estimators are derived, and their finite sample properties are explored through extensive simulations. The practical use of the proposed model is further illustrated through two real data applications.

\end{abstract}


\section{Introduction}
\label{sec:Introduction}

There are several applications which involve modeling multivariate count data. For example, non-life insurance claim counts are typically multi-dimensional, as standard policies often include different types of coverage and any claim occurrence may generate several different types of losses. An accurate analysis of such multivariate data requires both suitable marginal and joint specifications. Marginally, counts may exhibit a surplus of zeros (e.g., when no claims occur) that must be adequately accounted for. This phenomena of zero inflation is common when dealing with insurance claim counts, see, e.g., \cite{yip2005modeling}, \cite{boucher2007risk}, \cite{sarul2015application}. While assuming independence among the components of multi-dimensional data simplifies the statistical analysis, such an assumption is generally unrealistic and can lead to inaccurate conclusions, see, e.g. \cite{genest2014modeling} or \cite{bermudez2011bayesian}. Indeed, capturing the joint behavior of a random vector of counts is critical in evaluating any form of global risk, see, e.g., \cite{Zhang2025} for a recent comparative analysis of various approaches in the specific setting of multi-dimensional automobile insurance claims.

When the marginal count variables display an excess of zeros, traditional models, such as the Poisson distribution, are no longer appropriate. Early contributions studying univariate zero-inflated data include \cite{cohen1960estimating}, who incorporated extra zeros in a Poisson model for applications in quality control. This idea was further used and developed in other settings, such as statistical process control, see, e.g., \cite{xie2001zero}. An overabundance of zeros occurs in many other fields, and various modeling approaches exist for handling such data; some examples include applications in biology (e.g., \cite{martin1965fitting}, \cite{potts2006comparing}), agriculture (e.g.,  \cite{ridout1998models}), and in healthcare and epidemiology (e.g., \cite{bohning1999zero}, \cite{baetschmann2013modeling}, \cite{mwalili2008zero}). In the specific context of insurance claim counts, \cite{boucher2007risk} provide a comparative review of several (univariate) models, including the zero-inflated mixed Poisson and hurdle models.

Extensions to the multivariate setting can be done in many different ways. In the field of quality control, for example, \cite{li1999multivariate} extended the model of \cite{cohen1960estimating} to the multivariate setting by considering a mixture between different masses at zero and a common-shock multivariate Poisson model. In its full generality, the multivariate zero-inflated Poisson model proposed by \cite{li1999multivariate} contains many parameters rendering its use somewhat complex. \cite{liu2015type} present a simpler model that assumes independence between the counts, but captures some dependence through a common inflation factor driving the excess of zeroes across all components. \cite{wu2023multivariate} increase the flexibility by allowing different rates of zeroes for all margins. As an alternative, the hurdle model approach offers more control on the rates of zeroes by leveraging truncation, notably allowing for both an over- and under-abundance of zeroes.  A hurdle model is considered by \cite{tian2018type} to extend the work of \cite{liu2015type}. They develop a class of zero-adjusted Poisson models by mixing and truncating independent Poisson variables with a common Bernoulli inflation factor, allowing for both zero-inflation and zero-deflation.   In further developments, \cite{zhang2022new} consider a general distribution for the marginal counts, as well as an individual control on the rate of zeroes for all margins from multiple Bernoulli variables.  \cite{zhang2015two}, on the other hand, proposed two classes of distributions for zero-inflated count data, building on the bivariate generalized Poisson model of \cite{famoye2010}, which leverage a multiplicative factor to achieve positive and negative correlations constrained within some interval. \cite{zhang2015two} extend this model to allow for an excess of zeros by including a random Bernoulli inflation factor: the type I version considers a single inflation factor whereas the type II uses a pair of independent Bernoulli factors.

 The present work contributes to the literature on multivariate models for zero-inflated counts. The proposed model considers a common inflation factor to account for excess zeros, along similar lines as \cite{liu2015type}, thereby capturing some form of association among the components. To allow for further flexibility in the underlying dependence structure, our proposed formulation considers both comonotonic and counter-monotonic shocks in the latent count variables, leveraging the bivariate Poisson distribution of \cite{genest2018new}. As will be shown, the proposed construction permits both positive and negative correlations, while retaining an interpretable data generating structure.
This article is organized as follows. Section \ref{sec:1.model} presents the proposed model formulation, distinguishing between the case of positive and negative dependence. The implied dependence structure is then detailed in Section \ref{sec:2.dependence}. Section \ref{sec:3.Estimation} discusses various estimation approaches, including both moment-based and likelihood-based techniques. The model is studied through several simulations, summarized in Section \ref{sec:simulations}. A real data illustration involving catastrophic events is then provided in Section \ref{sec:data}. A brief discussion of higher dimensional settings is provided in Section \ref{sec: multivariateEX}, and finally concluding remarks are given in Section \ref{sec:conclusion}.

\section{Proposed model}
\label{sec:1.model}

The proposed class of bivariate zero-inflated Poisson models is based on a common inflation factor, acting on both margins. In particular, a pair $\mathbf{X}=(X_1,X_2)$ of correlated zero-inflated Poisson random variables are generated according to
\begin{equation}
\label{equ:bzip}
    \mathbf{X} = W \mathbf{T}
\end{equation}
where $W$ is a Bernoulli random variable with mean $1-\phi$, which is independent of the pair $\mathbf{T}=(T_1,T_2)$ of Poisson-distributed random variables with respective rates $\lambda_1$ and $\lambda_2$. Note that the construction in \eqref{equ:bzip} ensures that both components $T_1$ and $T_2$ consist of univariate zero-inflated Poisson random variables, see, e.g., \cite{lambert1992zero} and \cite{johnson/kotz/kemp}, with a common inflation parameter $\phi$ and respective counting rates $\lambda_1$ and $\lambda_2$. That is, for $j=1,2$, $X_j \sim \mathcal{ZIP}(\lambda_j,\phi)$ with
\[
P(X_j=x)=
\begin{cases}
    \phi + (1-\phi) e^{-\lambda_j} &\quad \text{for } x=0 \\
    (1-\phi) e^{-\lambda_j} \lambda_j^x / x! &\quad \text{for } x>0.
\end{cases}
\]
Thus, marginally, each component $X_j$ has mean $\E(X_j) = (1-\phi)\lambda_j$ and variance $\var(X_j) = (1-\phi)\lambda_j(1+\phi \lambda_j)$, $j=1,2$. 

It is clear that the common inflation factor will induce a form of association amongst the components of $(X_1,X_2)$, and the joint distribution of the latent counting process $(T_1,T_2)$ will further characterize the joint behavior of the pair $(X_1,X_2)$. While \cite{liu2015type} consider the same construction given in \eqref{equ:bzip}, the latter assumes independence between the counting components $T_1$ and $T_2$. In this work, we extend this formulation by allowing $(T_1,T_2)$ to follow the comonotonic or counter-monotonic shock bivariate Poisson models of \cite{genest2018new}. 

The class of bivariate Poisson models developed by \cite{genest2018new} generate correlated Poisson variables via convolutions of independent components with comonotonic or counter-monotonic shocks. More specifically, \cite{genest2018new} define a pair of dependent Poisson random variables according to $(T_1,T_2)=(Y_1+Z_1,Y_2+Z_2)$ where the latent variables $(Y_1,Y_2)$ are independent Poisson random variables, with respective marginal rates $(1-\theta)\lambda_1$ and $(1-\theta)\lambda_2$, which are independent of the shock variables $(Z_1,Z_2)$. In the model for positive dependence, the latent shock $(Z_1,Z_2)$ consists of a comonotonic pair with each component following a Poisson distribution with respective rates $\theta \lambda_1$ and $\theta \lambda_2$. That is,
\[
(Z_1,Z_2) = \left\{ G_{\theta \lambda_1}^{-1}(U),G_{\theta \lambda_2}^{-1}{(U)} \right\}
\]
where $U \sim \mathcal{U}(0,1)$ is a standard uniform random variable on the interval $(0,1)$, and $G_{\mu}$ is used to denote the univariate distribution function of a Poisson random variable with mean $\mu$. Negative dependence ensues when $(Z_1,Z_2)$ is defined as a counter-monotonic Poisson pair, again with respective rates $\theta \lambda_1$ and $\theta \lambda_2$, so that 
\[
(Z_1,Z_2) = \left\{ G_{\theta \lambda_1}^{-1}(U),G_{\theta \lambda_2}^{-1}(1-U) \right\}\]
where, once again, the pair is generated in terms of a common uniform random variable $U \sim \mathcal{U}(0,1)$. For a discussion of comonotonicity and counter-monotonicity, particularly in the case of Poisson margins, see \cite{Schulz2025}.

As shown in \cite{genest2018new}, the class of bivariate Poisson models based on comonotonic and counter-monotonic shocks allows for full flexibility in characterizing varying degrees of dependence, ranging from perfect negative dependence (when $\theta=1$ in the counter-monotonic shock model) to perfect positive dependence (when $\theta=1$ in the comonotonic shock model). Incorporating this class of bivariate Poisson models in the construction \eqref{equ:bzip} will thus also allow for further flexibility in capturing different strengths of dependence. 

In what follows, details on the proposed class of bivariate zero-inflated Poisson models will be provided, distinguishing from the case of positive and negative dependence. Throughout the text, the bivariate comonotonic shock Poisson model of \cite{genest2018new}  will be denoted as $\mathcal{BP}^+(\Lambda,\theta)$, and the counter-monotonic shock model will be written as $\mathcal{BP}^-(\Lambda,\theta)$, where $\Lambda = (\lambda_1,\lambda_2)$. The notation $\mathcal{P}(\lambda)$ will also be used to represent the univariate Poisson distribution with rate $\lambda$, while $g_{\lambda}$, $G_{\lambda}$ and $\bar{G}_{\lambda}$ will respectively denote the corresponding univariate probability mass function, distribution function and survival function.

\subsection{Positive dependence}

The model for positive dependence is based on the comonotonic shock formulation for the count components. Indeed, assuming $(T_1,T_2)\sim \mathcal{BP}^+(\Lambda,\theta)$ in \eqref{equ:bzip} induces a pair $(X_1,X_2)$ of positively correlated zero-inflated Poisson random variables, which will be denoted as $(X_1,X_2) \sim \mathcal{BZIP}^+(\Lambda,\theta,\phi)$. The joint probability mass function of $(X_1,X_2)$, expressed as $h^+_{\Lambda,\phi,\theta}$, can be obtained by conditioning on the latent inflation factor $W$, leading to
\begin{equation}\label{jointpmf}
    h^+_{\Lambda,\phi,\theta}(x_1,x_2) = \left\{\phi+(1-\phi) f^+_{\Lambda,\theta}(0,0)\right\}^{\delta_1 \delta_2} \times \left\{(1-\phi)f^+_{\Lambda,\theta}(x_1,x_2)\right\}^{1-\delta_1 \delta_2}
\end{equation}
where $\delta_j=\mathbf{1}(x_j=0)$ for $j=1,2$, with $\mathbf{1}(\cdot)$ representing the indicator function, and $f^+_{\Lambda,\theta}(t_1,t_2)=P(T_1=t_1,T_2=t_2)$, as given in \cite{Schulz/Genest/Mesfioui:2021}. The latter can be obtained by conditioning on the latent comonotonic shock $(Z_1,Z_2)$ viz.
\[
f^+_{\Lambda,\theta}(t_1,t_2)=\sum_{z_1=0}^{t_1} \sum_{z_2=0}^{t_2} g_{(1-\theta)\lambda_1}(t_1-z_1) g_{(1-\theta)\lambda_2}(t_2-z_2)c_{\Lambda,\theta}^+(z_1,z_2) 
\]
where $c_{\Lambda,\theta}^+(z_1,z_2)$ is the joint probability mass function of the comonotonic pair $(Z_1,Z_2)$ given by
\[
c_{\Lambda,\theta}^+(z_1,z_2) = \left[ \min\left\{ G_{\theta\lambda_1}(z_1),G_{\theta\lambda_2}(z_2)\right\} - \max\left\{ G_{\theta\lambda_1}(z_1-1), G_{\theta\lambda_2}(z_2-1) \right\}\right]_+
\]
with the notation $[x]_+ = x \mathbf{1}(x>0)$. 

Similarly, by conditioning on the Bernoulli latent variable $W$, one obtains an expression for the joint distribution function, $H^+_{\Lambda,\phi,\theta}$, which can be written as follows
\begin{equation}\label{jointcdf}
    H^+_{\Lambda,\phi,\theta}(x_1,x_2) = \phi \Delta_1\Delta_2 + (1-\phi) F^+_{\Lambda,\theta}(x_1,x_2),
\end{equation}
where $\Delta_j = \mathbf{1}(x_j \geq0)$ and $F^+_{\Lambda,\theta}$ is the joint distribution of $(T_1,T_2) \sim \mathcal{BP}^+(\Lambda,\theta)$. The latter is again obtained by conditioning on the latent comonotonic shock $(Z_1,Z_2)$, as shown in \cite{genest2018new}, and can be written as
\[
F^+_{\Lambda,\theta}(t_1,t_2) = \sum_{z_1=0}^{t_1} \sum_{z_2=0}^{t_2} G_{(1-\theta)\lambda_1}(t_1-z_1) G_{(1-\theta)\lambda_2}(t_2-z_2)c_{\Lambda,\theta}^+(z_1,z_2) .
\]

Other distributional properties of the proposed $\mathcal{BZIP^+}$ class can be derived in a similar manner, by conditioning on the latent zero-inflation factor $W$. For example, the probability generating function, $G^+_{\Lambda,\phi,\theta}=\E(s_1^{X_1} s_2^{X_2})$ is obtained as 
\begin{equation*} \label{eqPGF}
G^+_{\Lambda,\phi,\theta}(s_1,s_2) = \phi + (1-\phi)\varrho^+_{\Lambda,\theta}(s_1,s_2) \exp\left\{(1-\theta)\lambda_1(s_1-1) + (1-\theta)\lambda_2(s_2-1)\right\},
\end{equation*}
where $\varrho^+_{\Lambda,\theta}(s_1,s_2) = E({s_1}^{Z_1}{s_2}^{Z_2})$ as given in \cite{genest2018new}.

\subsection{Negative dependence}

Now suppose $(T_1,T_2)\sim \mathcal{BP}^-(\Lambda,\theta)$ in \eqref{equ:bzip}, so that the underlying latent counts in the construction are negatively correlated. In this setting, the pair $(X_1,X_2) \sim \mathcal{BZIP}^-(\Lambda,\theta,\phi)$. Similarly to the case of positive dependence, expressions for the joint probability and distribution functions of $(X_1,X_2)$, respectively denoted by $h^-_{\Lambda,\phi,\theta}$ and $H^-_{\Lambda,\phi,\theta}$, follow directly from the joint behavior of $(T_1,T_2)$. In particular, the joint PMF is given by
\begin{equation}\label{jointpmf_neg}
    h^-_{\Lambda,\phi,\theta}(x_1,x_2) = \left\{\phi+(1-\phi) f^-_{\Lambda,\theta}(0,0)\right\}^{\delta_1 \delta_2} \times \left\{(1-\phi)f^-_{\Lambda,\theta}(x_1,x_2)\right\}^{1-\delta_1 \delta_2} \\
\end{equation}
and the joint CDF is
\begin{equation}\label{jointcdf_neg}
    H^-_{\Lambda,\phi,\theta}(x_1,x_2) = \phi\Delta_1\Delta_2 + (1-\phi) F^-_{\Lambda,\theta}(x_1,x_2),
\end{equation}
where, once again, $f^-_{\Lambda,\theta}$ and $F^-_{\Lambda,\theta}$ represent the joint PMF and CDF of $(T_1,T_2)\sim \mathcal{BP}^-(\Lambda,\theta)$, respectively given by
\begin{align*}
f^-_{\Lambda,\theta}(t_1,t_2) & =\sum_{z_1=0}^{t_1} \sum_{z_2=0}^{t_2} g_{(1-\theta)\lambda_1}(t_1-z_1) g_{(1-\theta)\lambda_2}(t_2-z_2)c^-_{\Lambda,\theta}(z_1,z_2) \\
F^-_{\Lambda,\theta}(t_1,t_2) & =\sum_{z_1=0}^{t_1} \sum_{z_2=0}^{t_2} G_{(1-\theta)\lambda_1}(t_1-z_1) G_{(1-\theta)\lambda_2}(t_2-z_2)c^-_{\Lambda,\theta}(z_1,z_2) 
\end{align*}
In the above expressions, $c^-_{\Lambda,\theta}(z_1,z_2) $ is now used to denote the joint probability mass function of the counter-monotonic pair $(Z_1,Z_2)$, which can be shown to be given by
\[
c^-_{\Lambda,\theta}(z_1,z_2) = \left[ \min\left\{ G_{\theta\lambda_1}(z_1),\bar{G}_{\theta\lambda_2}(z_2-1)\right\} - \max\left\{ G_{\theta\lambda_1}(z_1-1), \bar{G}_{\theta\lambda_2}(z_2) \right\}\right]_+ .
\]

The probability generating function, $G^-_{\Lambda,\phi,\theta}$ also leads to a similar expression to that obtained in the model for positive dependence, viz.
\begin{equation*} \label{eqPGFneg}
G^-_{\Lambda,\phi,\theta}(s_1,s_2) = \phi + (1-\phi)\varrho^-_{\Lambda,\theta}(s_1,s_2) \exp\left\{(1-\theta)\lambda_1(s_1-1) + (1-\theta)\lambda_2(s_2-1)\right\},
\end{equation*}
where $\varrho^-_{\Lambda,\theta}(s_1,s_2)$ is the probability generating function of the counter-monotonic shock $(Z_1,Z_2)$, see \cite{genest2018new}.

\section{Dependence structure} \label{sec:2.dependence}

In the proposed model given in \eqref{equ:bzip}, the dependence between the zero-inflated counts $(X_1,X_2)$ is induced not only from the common inflation factor $W$, but also from the joint behavior of the latent counting pair $(T_1,T_2)$. Accordingly, there are two parameters regulating the strength of dependence: $\phi$, which controls the rate of the inflation, and $\theta$, which controls the degree of dependence between the latent count variables. Indeed, the covariance for a pair $(X_1,X_2)\sim \mathcal{BZIP}(\Lambda,\theta,\phi)$ is given by
\begin{equation}\label{equ:cov}
    \cov(X_1,X_2)=(1-\phi)\cov(T_1,T_2) + \phi(1-\phi)\lambda_1 \lambda_2,
\end{equation}
with corresponding correlation 
\begin{equation}\label{equ:corr}
    \corr(X_1,X_2)= \frac{\corr(T_1,T_2)}{\sqrt{(1+\phi \lambda_1)(1+\phi \lambda_2) }} + \frac{\phi \lambda_1 \lambda_2}{\sqrt{(1+\phi \lambda_1)(1+\phi \lambda_2) \lambda_1 \lambda_2}}.
\end{equation}

When $\theta = 0$, the counting pair $(T_1,T_2)$ is independent and $\corr(T_1,T_2)=0$. In this case, the only source of dependence is the common inflation factor $W$ and the correlation then simplifies to 
\[
\corr(X_1,X_2)=\phi \sqrt{\lambda_1 \lambda_2 / \{(1+\phi\lambda_1)(1+\phi\lambda_2)\}},
\]
which coincides with what is shown in \cite{liu2015type}. In the model for positive dependence with $(T_1,T_2) \sim \mathcal{BP}^+(\Lambda,\theta)$, \cite{genest2018new} show that $\corr(T_1,T_2)$ is an increasing function of $\theta$, given by
\[
\frac{1}{\sqrt{\lambda_1 \lambda_2}} \left[ -\theta^2 \lambda_1 \lambda_2 + \sum_{i=0}^{\infty} \sum_{j=0}^{\infty} \min \left\{ \bar{G}_{\theta \lambda_1}(i), \bar{G}_{\theta \lambda_2}(j) \right\} \right].
\]
In this case, the zero-inflated model will yield a correlation $\corr(X_1,X_2)$ which is also an increasing function of $\theta$. On the other hand, when $(T_1,T_2)$ stems from the counter-monotonic shock model, \cite{genest2018new} show that the correlation $\corr(T_1,T_2)$ is a decreasing function of $\theta$, which can be expressed as
\[
\frac{1}{\sqrt{\lambda_1 \lambda_2}} \left[ -\theta^2 \lambda_1 \lambda_2 + \sum_{i=0}^{\infty} \sum_{j=0}^{\infty} \min \left\{0, G_{\theta \lambda_1}(i) + G_{\theta \lambda_2}(j) -1 \right\} \right].
\]
Thus, in the model for negative dependence, the implied correlation $\corr(X_1,X_2)$ will also be a decreasing function of $\theta$.

In general, the effect of $\phi$ on the dependence is not straightforward as this parameter not only regulates a portion of the association between $X_1$ and $X_2$, but also the marginal behavior of each component. In the extremal case where $\phi=0$, $\corr(X_1,X_2)=\corr(T_1,T_2)$, and the model simplifies to that of the bivariate Poisson shock models of \cite{genest2018new}. On the other hand, when $\phi=1$, the correlation is undefined as $(X_1,X_2)$ are degenerate random variables which place all the mass at $(0,0)$. 

For fixed $\lambda_1>0$, $\lambda_2>0$, one can express the correlation given in \eqref{equ:corr} as a function of $\phi \in (0,1)$:
\begin{equation}\label{equ:corr_v2}
    \corr(X_1,X_2) = a_{\Lambda}(\phi) \corr(T_1,T_2)  + b_{\Lambda}(\phi)
\end{equation}
where $a_{\Lambda}(\phi) = \{(1+\phi \lambda_1)(1+\phi \lambda_2)\}^{-1/2}$ and $b_{\Lambda}(\phi) = \phi  \, a_{\Lambda}(\phi) \sqrt{\lambda_1 \lambda_2}$. It is clear that $a_{\Lambda}(\phi)$ is a decreasing function of $\phi$. Accordingly, when $(T_1,T_2)\sim \mathcal{BP}^+(\Lambda,\theta)$, one has that $a_{\Lambda}(\phi) \corr(T_1,T_2)$ is  decreasing in $\phi$, whereas for $(T_1,T_2)\sim \mathcal{BP}^-(\Lambda,\theta)$ it follows that $a_{\Lambda}(\phi) \corr(T_1,T_2)$ is increasing in $\phi$. It can be shown that the function $b_{\Lambda}(\phi)$ in \eqref{equ:corr_v2} is increasing in $\phi$, since for fixed $\Lambda \in (0,\infty)^2$, $\partial b_{\Lambda}(\phi)/\partial \phi > 0$ for any $\phi \in (0,1)$. 

\begin{figure}[htbp]
    \centering
    \includegraphics[width= 0.6\textwidth]{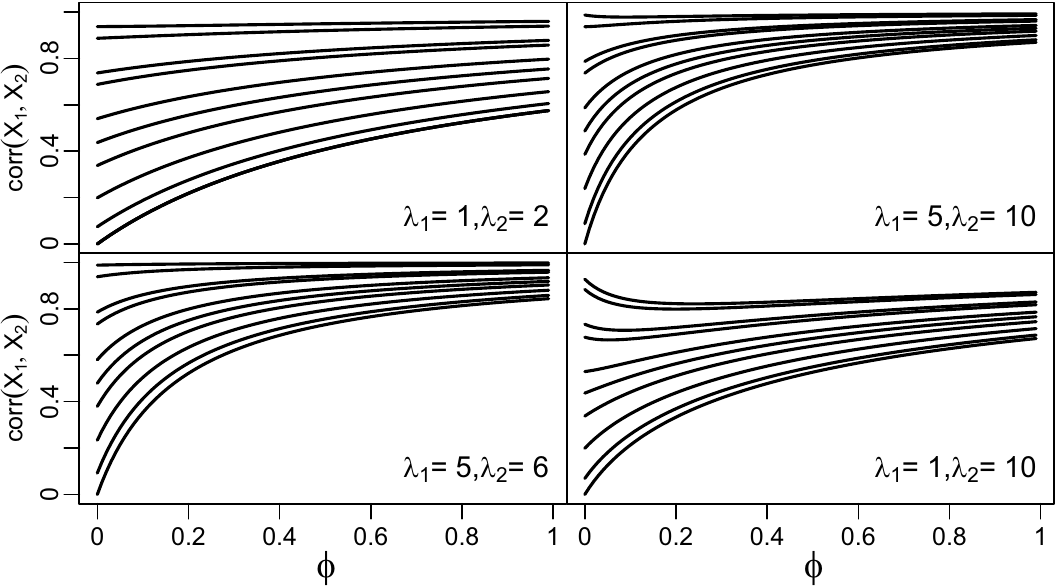}
        \label{fig:plot2}
       \caption{Correlation as a function of $\phi$ in the model for positive dependence. In each plot, the distinct lines represent different values of $\theta$ ranging, from bottom to top, in $\{0.1, 0.25, 0.4, 0.5, 0.6, 0.75, 0.8, 0.95\}$. The four plots depict the relation for varying values of $(\lambda_1,\lambda_2)$ with $(1,2)$ shown in the top left panel, $(5,10)$ in the top right, $(5,6)$ bottom left, and $(1,10)$ bottom right.
}
    \label{fig:corr_phi_positive}
\end{figure}

\begin{figure}[htbp]
    \centering
        \includegraphics[width= 0.6\textwidth]{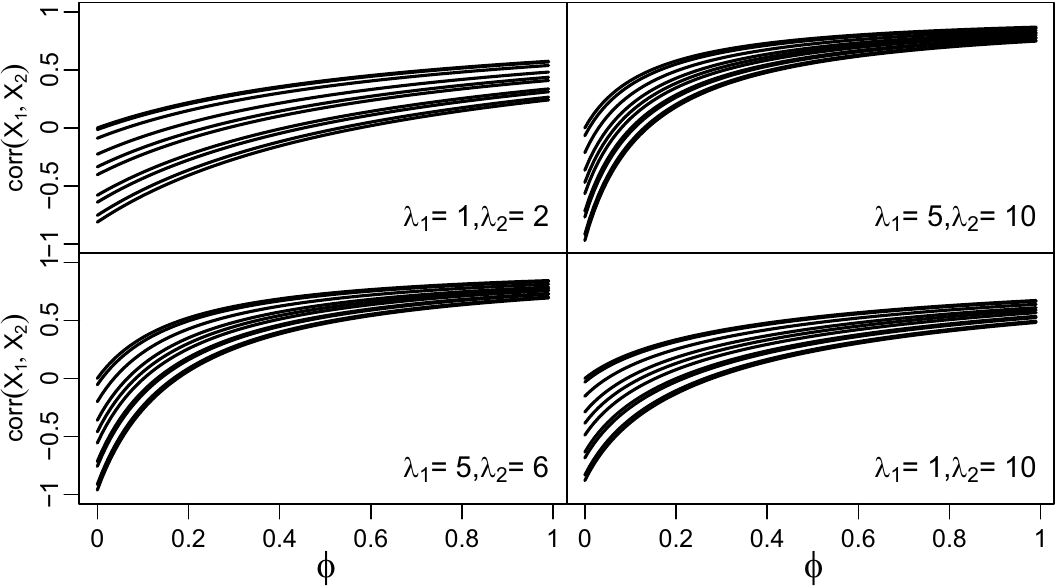}
        \label{fig:plot1}
        \caption{Correlation as a function of $\phi$ in the model for negative dependence. In each plot, the distinct lines represent different values of $\theta$ ranging, from bottom to top, in $\{0.1, 0.25, 0.4, 0.5, 0.6, 0.75, 0.8, 0.95\}$ . The four plots depict the relation for varying values of $(\lambda_1,\lambda_2)$ with $(1,2)$ shown in the top left panel, $(5,10)$ in the top right, $(5,6)$ bottom left, and $(1,10)$ bottom right.
    }
    \label{fig:corr_phi_negative}
\end{figure}

It thus follows that in the model for negative dependence, fixing $\Lambda$ and $\theta$, the pairwise correlation $\corr(X_1,X_2)$ is always an increasing function of $\phi$. This is intuitive as increasing the zero-inflation rate correspondingly increases the probability of common zeros for both components, thereby pulling towards a form of positive association. On the other hand, in the model for positive dependence, $\corr(X_1,X_2)$ can be both increasing and decreasing in $\phi$, depending on the values of $\Lambda$ and $\theta$. This is depicted in Figures~\ref{fig:corr_phi_positive} and \ref{fig:corr_phi_negative}, in the models for positive and negative dependence, respectively. 

In certain specific cases, further simplifications are possible. For example, in the model for positive dependence, setting $\lambda_1 = \lambda_2 = \lambda$ leads to 
\[
\corr(X_1,X_2) = \frac{\theta + \phi\lambda}{1+\phi\lambda}.
\]
This follows directly since when $\lambda_1=\lambda_2$, the counting pair $(T_1,T_2)$ follows the common shock model viz. $(T_1,T_2)=(Y_1+Z,Y_2+Z)$, where the latent variables $Y_1, Y_2 \overset{iid}{\sim}\mathcal{P}\left\{(1-\theta)\lambda\right\}$, and are independent of the common shock $Z \mathcal \sim {P}(\theta \lambda)$. For further details on the common shock bivariate Poisson model, see, e.g., \cite{campbell1934poisson} and \cite{holgate1964estimation}. Further, in this setting, when $\theta=1$, $\corr(X_1,X_2)=1$ for any $\phi \in (0,1)$ as in this case $X_1=X_2=WZ$.

\subsection{Strength of association}

As previously explored, the parameter $\theta$ regulates the strength of the dependence stemming from the counting process, thereby controlling the degree of association between $(X_1,X_2)$. Indeed, the proposed class of bivariate zero-inflated Poisson models are ordered in the positive quadrant dependent (PQD) ordering. This stems directly from the PQD ordering property of the class of $\mathcal{BP}$ models, see \cite{genest2018new}. The latter established that in the comonotonic shock model, $\theta < \theta^{\prime} \Rightarrow F^+_{\Lambda,\theta}(i,j) \leq F^+_{\Lambda,\theta^{\prime}}(i,j)$ for any $i,j \in \mathbb{N}$ and fixed $\Lambda \in (0,\infty)^2$. Analogously, in the counter-monotonic shock model, fixing $\Lambda \in (0,\infty)^2$, one has that $\theta < \theta^{\prime} \Rightarrow F^-_{\Lambda,\theta}(i,j) \geq F^-_{\Lambda,\theta^{\prime}}(i,j)$ for any $i,j \in \mathbb{N}$. This then directly extends to the proposed class of bivariate zero-inflated Poisson models based on the forms for the joint cumulative distribution functions \eqref{jointcdf} and \eqref{jointcdf_neg}. The PQD ordering property of the proposed $\mathcal{BZIP}$ class is formalized in the following two corollaries. 

\begin{corollary}
Suppose $(X_1,X_2) \sim \mathcal{BZIP}^+(\Lambda,\theta,\phi)$ and $(X_1^{\prime},X_2^{\prime}) \sim \mathcal{BZIP}^+(\Lambda,\theta^{\prime},\phi)$. Then $\theta < \theta^{\prime} \Rightarrow (X_1,X_2) \prec_{PQD} (X_1^{\prime},X_2^{\prime})$.
\end{corollary}

\begin{corollary}
Suppose $(X_1,X_2) \sim \mathcal{BZIP}^-(\Lambda,\theta,\phi)$ and $(X_1^{\prime},X_2^{\prime}) \sim \mathcal{BZIP}^-(\Lambda,\theta^{\prime},\phi)$. Then $\theta < \theta^{\prime} \Rightarrow (X_1^{\prime},X_2^{\prime}) \prec_{PQD} (X_1,X_2) $.
\end{corollary}

As is well known, the joint distribution $H$ of a random pair $(X_1,X_2)$ with given marginal CDFs, $F_1$ and $F_2$, satisfies
\[
\max\left\{ 0,F_1(x_1)+F_2(x_2)-1 \right\} \leq H(x_1,x_2) \leq \min \left\{ F_1(x_1),F_2(x_2) \right\}.
\]
The upper and lower bounds in the above inequality respectively represent the upper and lower Fréchet-Hoeffding bounds, see, e.g., Section~2.5 of \cite{Nelsen:textbook}. In the case where the margins consist of zero-inflated Poisson random variables with a common inflation rate, i.e., $X_j \sim \mathcal{ZIP}(\phi,\lambda_j)$, $j=1,2$, the Fréchet-Hoeffding upper bound simplifies to
\[
\phi + (1-\phi) \min \left\{ G_{\lambda_1}(x_1),G_{\lambda_2}(x_2) \right\}.
\]
It can be shown that $F^+_{\Lambda,\theta=1}(x_1,x_2)=\min \left\{ G_{\lambda_1}(x_1),G_{\lambda_2}(x_2) \right\}$, and thus $H^+_{\Lambda,\phi,\theta}(x_1,x_2)$ attains the upper Fréchet-Hoeffding bound when $\theta=1$. That is, $(X_1,X_2) \sim \mathcal{BZIP}^+(\Lambda,\theta,\phi)$ attains the strongest degree of positive dependence possible for an arbitrary pair of zero-inflated Poisson random variables of the form $X_j \sim \mathcal{ZIP}(\phi,\lambda_j)$, $j=1,2$, whenever $\theta=1$. As such, for $\theta=1$, $(X_1,X_2) \sim \mathcal{BZIP}^+(\Lambda,\theta,\phi)$ consist of a comonotonic pair. 

Note that the same does not follow in the case of negative dependence. Indeed, the lower Fréchet-Hoeffding bound on the joint distribution for a pair $(X_1,X_2)$ with marginal specifications $X_j \sim \mathcal{ZIP}(\phi,\lambda_j)$, $j=1,2$, is given by
\[
\max\Big[0,\phi + (1-\phi)\left\{ G_{\lambda_1}(x_1)+G_{\lambda_2}(x_2)-1 \right\} \Big].
\]
In the extremal case where $\theta=1$, it can be shown that the $\mathcal{BZIP}^-(\Lambda,\theta,\phi)$ model has joint distribution given by $H^-_{\Lambda,\phi,\theta}(x_1,x_2)=\phi + (1-\phi) \max \left\{0,G_{\lambda_1}+G_{\lambda_2}-1\right\}$. The latter only coincides with the lower Fréchet-Hoeffding bound when it is further imposed that $\phi=0$, that is, the case where there is no zero-inflation and the model simplifies to the bivariate counter-monotonic shock Poisson model of \cite{genest2018new}.

\section{Estimation}\label{sec:3.Estimation}

Let $\mathbf{X} = \{(X_{11},X_{12}), \ldots, (X_{n1},X_{n2})\} $, denote a random sample from the proposed $\mathcal{BZIP}$ family. It is of interest to estimate the model parameters $\Psi=(\lambda_1,\lambda_2,\theta,\phi)$, where $(\lambda_1,\lambda_2) \in (0,\infty)^2$ and each of $\theta$ and $\phi$ fall in the interval $(0,1)$. To this end, there are various approaches possible, including both moment-based and likelihood-based methods. In what follows, both estimation methods will be detailed, as well as an iterative procedure for obtaining the maximum likelihood estimators based on the EM algorithm. 

\subsection{Method of moments}\label{sec:MM}

As is well known, the method of moments (MoM) relies on matching theoretical moments to the empirical moments of a random sample. In the case of the $\mathcal{BZIP}$ class of distributions, the marginal parameters $\Lambda \in (0,\infty)^2$, as well as the common inflation rate $\phi \in (0,1)$, can be estimated via the first two marginal moments. 

For $X_j \sim \mathcal{ZIP}(\phi_j,\lambda_j)$, $j = 1, 2$, the first and second moments are respectively given by $(1-\phi_j)\lambda_j$ and $(1-\phi_j)\lambda_j(1+\lambda_j)$. The MoM estimators, $\check{\lambda}_j$ and $\check{\phi}_j$, are then obtained by solving the following system of equations
\begin{equation*}
    \begin{split}
       (1-\phi)\lambda_j &= \bar{X}_j\\
    (1-\phi)\lambda_j(1+\lambda_j) &= \frac{1}{n}\sum_{i = 1}^{n}{X_{ij}^2},
    \end{split}
\end{equation*}
which yields 
\begin{equation*}
   \check{\lambda}_j = \frac{\sum_{i=1}^{n}{X_{ij}^2}}{n\bar{X_j}}-1, \quad \quad
        \check{\phi}_j = 1-\frac{\bar{X}_j}{\check{\lambda}_j} .
\end{equation*}
The proposed class of bivariate zero-inflated Poisson models assumes a common inflation factor for both margins. Thus, to obtain a single estimator one can set $\check{\phi} = (\check{\phi}_1+ \check{\phi}_2)/2$. 

Estimation of the dependence parameter $\theta$ stems from the pairwise covariance, given in \eqref{equ:cov}. Letting $s_{\Lambda,\phi}(\theta)=\cov(X_1, X_2)$, the MoM estimator $\check{\theta}$ is then the root of $s_{\check{\Lambda},\check{\phi}}(\theta) - S_{12}$, where $S_{12}$ denotes the sample covariance $\sum_{i=1}^n (X_{i1}-\bar{X}_1)(X_{i2}-\bar{X}_2)/(n-1)$. Recall from Section~\ref{sec:2.dependence} that for fixed $\Lambda$ and $\phi$, the covariance $s_{\Lambda,\phi}(\theta)$ is a monotonic function of $\theta$ and thus a unique root exists. Standard statistical results allow to ascertain that the MoM estimators $\check{\Psi}=(\check{\Lambda},\check{\theta},\check{\phi})$ are consistent and asymptotically normally distributed, as formalized in the following theorem; see Appendix~\ref{app:MM} for a detailed proof.

\begin{theorem}
Suppose that $(X_{i1},X_{i2}) \overset{iid}{\sim} \mathcal{BZIP}(\Lambda,\theta,\phi)$, for $i=1,\ldots,n$. Then, as $n \rightarrow \infty$,
\begin{align*}
    \sqrt{n}(\check{\lambda}_j - \lambda_j) & \rightsquigarrow \mathcal{N}(0, \xi^2) \\
    \sqrt{n}(\check{\phi}-\phi) & \rightsquigarrow \mathcal{N}(0,\zeta^2) \\
    \sqrt{n}(\check{\theta}-\theta) &\rightsquigarrow \mathcal{N}(0,\epsilon^2) \\
\end{align*}
where $\rightsquigarrow$ denotes convergence in distribution, 
\[
\begin{aligned}
    \xi^2 &= \sigma / \{(1+\phi)\lambda_j\}, \\
    \zeta^2 &= \frac{1-\phi}{4}\left\{ \frac{1+\phi\lambda_1}{\lambda_1}+ \frac{1+\phi\lambda_2}{\lambda_2} \right\} + \frac{n}{2}\cov(\check{\phi}_1,\check{\phi}_2), \\
    \epsilon^2 &= \left\{\gamma^{\prime}(\sigma_{12})\right\}^2 \tau^2 ,\\
\end{aligned}
\]
with $\gamma$ denoting the inverse covariance function, i.e. $\gamma : \theta \mapsto s^{-1}_{\Lambda,\phi}(\theta)$, $\gamma^{\prime}$ denoting the corresponding derivative, $\sigma_{12}=\cov(X_1,X_2)$ and $\tau^2 = \var\left[\{X_1-(1-\phi)\lambda_1\}\{X_2-(1-\phi)\lambda_2\} \right]$.
\end{theorem}

Note that the above estimation procedure does not guarantee that the parameter constraints will necessarily be respected, and in some instances further caution is required. 
For example, if one, or both, of the components in the sample take on only the values $0$ or $1$, i.e., $X_{ij} \in \{0,1\} $ for each $i\in \{1,\ldots,n\}$, the estimate $\check{\lambda}_j$ will either be $0$ or undefined, and consequently $\check{\phi}_j$ will be undefined. While such settings are unlikely, when $\phi$ is large and $\lambda_j$ is small, this can indeed occur with non-negligible probability, particularly in small samples. This will be further addressed in the simulation study in Section~\ref{sec:simulations}. 

Whenever the method of moments results in reasonable estimates for $\Lambda$ and $\phi$, one can proceed to estimating the dependence parameter $\theta$. As previously explored, for fixed $\Lambda \in (0,\infty)^2$ and $\phi \in (0,1)$, the implied covariance in the model for positive dependence, denoted as $s_{\Lambda,\phi}^+(\theta)$, is increasing in $\theta$. When setting the parameters $\Lambda$ and $\phi$ equal to their respective MoM estimates, it could happen that the observed sample covariance extends beyond the permissible range implied by the model, i.e., $S_{12}>s_{\check{\Lambda},\check{\phi}}^+(1)$ or $ S_{12}<s_{\check{\Lambda},\check{\phi}}^+(0)$. In this case, one can constrain $\check{\theta}$ to fall within the interval $(0,1)$ by capping either at $0$ or $1$, as the case may be. Analogously, the model for negative dependence results in a covariance, denoted as $s_{\Lambda,\phi}^-(\theta)$, which decreases in $\theta$. In a similar way, whenever the MoM estimates $\check{\Lambda}$ and $\check{\phi}$ result in $s_{\check{\Lambda},\check{\phi}}^-(1)<S_{12}$ or $s_{\check{\Lambda},\check{\phi}}^-(0)>S_{12}$, $\check{\theta}$ will be constrained accordingly to $1$ or $0$, as the case may be. Note that while the $\mathcal{BZIP}^+$ model always implies a positive covariance, the $\mathcal{BZIP}^-$ model does not necessarily entail a negative association between the components, as previously illustrated in Figure~\ref{fig:corr_phi_negative}. Rather, it is the portion of the covariance due to the latent count variables, $\cov(T_1,T_2)$, which reflects the positive and negative dependence settings. Let $S_{12}^*$ denote the portion of the sample covariance due to the latent count variables, that is, $S_{12}^* = S_{12}/(1-\phi) - \phi \lambda_1 \lambda_2$. In terms of method of moments estimation, one then requires that $S_{12}^*>0$ in the model for positive dependence, and $S_{12}^*<0$ in the model for negative dependence. The sample covariance is a consistent estimator for $\cov(X_1,X_2)$ and thus should respect these constraints. However, in finite samples such issues could arise; this will be further detailed in Section~\ref{sec:simulations}.   

\subsection{Maximum likelihood estimation}\label{sec:ML}

The maximum likelihood estimators, which will be denoted as $\hat{\Psi}$, are found by maximizing the joint likelihood $L(\Psi;\mathbf{X})$, or equivalently the log-likelihood $\ell(\Psi;\mathbf{X})$. 
An expression for the likelihood function is obtained directly from \eqref{jointpmf} or \eqref{jointpmf_neg}, according to the model for positive and negative dependence, as the case may be. In particular, define the set $M_0 = \{i \in \{1, \dots, n\}: (X_{i1},X_{i2}) = (0,0)\}$, and let $m_0$ consist of the number of observed joint zeros in the sample, i.e., $m_0 = | M_0 |$. In what follows, a unifying notation for the probability mass function will be used, with $f_{\Lambda,\theta}$ denoting the PMF stemming from either the positive or negative dependence model. The likelihood for $\mathbf{X}=\mathbf{x}$ is then given by
\[
L(\psi ; \mathbf{x}) 
        = \Big[\phi+(1-\phi)f_{\Lambda, \theta}(0,0)\Big]^{m_0} (1-\phi)^{(n-m_0)} \prod_{i \notin M_0}{f_{\Lambda,\theta}(x_{i1},x_{i2})}
\]
with corresponding log-likelihood 
\begin{equation*}\label{loglik}
         \ell (\psi ; \mathbf{x}) =  m_0 \log\Big[\phi+(1-\phi)f_{\Lambda, \theta}(0,0)\Big] 
         + (n-m_0) \log(1-\phi) + 
         \sum_{i \notin M_0}\log f_{\theta,\Lambda}(x_{i1},x_{i2})
\end{equation*}

It is straightforward to show that solving ${\partial \ell (\Psi;\mathbf{X}) }/{\partial \phi} = 0$ yields 
\begin{equation}\label{equ:phiest}
    \hat{\phi} = \frac{(\frac{m_0}{n}) - f_{\Lambda, \theta}(0,0)}{1 - f_{\Lambda, \theta}(0,0)}
\end{equation}
We remark that the form of $\hat{\phi}$ given in \eqref{equ:phiest} is quite intuitive. Indeed, the numerator in \eqref{equ:phiest} consists of the difference between the empirical probability of observing $(X_1,X_2)=(0,0)$, i.e., $m_0/n$, and the probability that the observed pair of zeros stems from the counting variables, i.e., $(T_1,T_2)=(0,0)$. The denominator of the estimator given in \eqref{equ:phiest} rescales by the probability that the counting variables are positive, i.e., $1-\Prob(T_1=0,T_2=0)$. And indeed, from both \eqref{jointpmf} and \eqref{jointpmf_neg}, one can see that $P\left\{(X_1,X_2)=(0,0)\right\}=\phi\left\{1-f_{\Lambda,\theta}(0,0) \right\} + f_{\Lambda,\theta}(0,0)$.

Substituting $\hat{\phi}$ into the log-likelihood leads to
\begin{equation}\label{loglik2}
    \begin{split}
        {\ell^*}(\hat{\phi}, \theta, \Lambda; \mathbf{x}) = m_0 \log \left\{\frac{m_0}{n-m_0} \right\} + n \log \left\{\frac{n - m_0}{n} \right\} + \sum_{i \notin M_0}{\log \left\{\frac{f_{\Lambda,\theta}(x_{i1},x_{i2})}{1-f_{\Lambda,\theta}(0,0)} \right\}}
    \end{split}
\end{equation}
Estimation of the remaining parameters $(\Lambda,\theta)$ can then be carried out based on \eqref{loglik2}, viz.
\[
(\hat{\Lambda},\hat{\theta}) = \argmax_{\Lambda, \theta} \sum_{i \notin M_0}{\log \left\{\frac{f_{\Lambda,\theta}(x_{i1},x_{i2})}{1-f_{\Lambda,\theta}(0,0)} \right\}},
\]
subject to the constraints $\Lambda \in (0,\infty)^2$ and $\theta \in (0,1)$. Note that a reparametrization can be considered by setting $\alpha_j = \log(\lambda_j)$, $j=1,2$, and $\pi=\log\{\theta/(1-\theta)\}$, thereby removing the parameter constraints in the optimization procedure. The maximum likelihood estimators (MLEs) are then $\hat{\lambda}_j=\exp(\hat{\alpha}_j)$, $j=1,2$, $\hat{\theta}=\exp(\hat\pi)/\{1+\exp(\hat\pi)\}$ and $\hat{\phi}$ is found by substituting the estimate $f_{\hat{\Lambda},\hat{\theta}}(0,0)$ into \eqref{equ:phiest}. 

Note that implementing this procedure could lead to issues in estimating $\hat{\phi}$ whenever $(m_0/n)<f_{\hat\Lambda, \hat \theta}(0,0)$, which may occur in smaller samples. An alternative implementation approach is to simultaneously optimize $\ell(\psi;\mathbf{x})$ across all parameters $\Psi$, using a similar reparameterization for $\phi$ as well to remove all parameter constraints. This will be further explored in simulations in Section~\ref{sec:simulations}.

Standard likelihood theory ensures that $\sqrt{n}(\hat{\Psi} -\Psi) \rightsquigarrow \mathcal{N}(\mathbf{0},\mathcal{I}^{-1} )$, where $\Psi$ denotes the true parameter values and $\mathcal{I}$ is the Fisher information matrix. Note that evaluating $\mathcal{I}$ in the proposed $\mathcal{BZIP}$ class is difficult, and in practice standard bootstrap methods could be used to estimate the asymptotic variance of $\hat{\Psi}$.

\subsection{EM algorithm}\label{sec:EM}
The proposed model formulation \eqref{equ:bzip} is based on the latent Bernoulli zero-inflation factor $W$, and as such, one could consider an EM algorithm approach for finding the maximum likelihood estimators. The augmented probability mass function has the form
\begin{equation}\label{equ:pmf_em}
    P(X_1=x_1,X_2=x_2,W=w)=\left\{ (1-\phi) f_{\Lambda,\theta}(x_1,x_2)\right\}^w \left\{ \phi \delta_1 \delta_2 \right\}^{1-w},
\end{equation}
where, as previously defined, $\delta_j=\mathbf{1}(x_j=0)$, $j=1,2$, and the unified notation $f_{\Lambda,\theta}(x_1,x_2)$ is used for $P(T_1=x_1,T_2=x_2)$ in either the positive or negative dependence setting. From this, it follows that
\begin{equation}\label{equ:EW_EM}
    \E(W|X_1=x_1,X_2=x_2) = \frac{(1-\phi)f_{\Lambda,\theta}(x_1,x_2)}{\left\{\phi+(1-\phi) f_{\Lambda,\theta}(0,0)\right\}^{\delta_1 \delta_2} \times \left\{(1-\phi)f_{\Lambda,\theta}(x_1,x_2)\right\}^{1-\delta_1 \delta_2}},
\end{equation}
which takes on the value $1$ whenever $(X_1,X_2) \neq (0,0)$. From \eqref{equ:pmf_em}, the complete data log-likelihood is given by
\[
\ell(\Psi;\mathbf{x},\mathbf{W}) = \sum_{i=1}^n \Bigr[ W_i \left\{ \log(1-\phi) + \log f_{\Lambda,\theta}(x_{i1},x_{i2}) -\log\phi -\log(\delta_{i1} \delta_{i2}) \right\} + \log \phi + \log(\delta_{i1} \delta_{i2}) \Bigr].
\]

Let $\hat{\Psi}^{(k)}$ denote the estimates obtained at the $k^{th}$ iteration of the EM algorithm. At the next iteration, the E-step then yields
\begin{multline}\label{equ:Q_em}
    Q(\Psi ; \Psi^{(k)}) = \sum_{i=1}^n \Bigr[ w(\Psi^{(k)};x_{i1},x_{i2}) \left\{ \log(1-\phi) + \log f_{\Lambda,\theta}(x_{i1},x_{i2}) -\log\phi -\log(\delta_{i1} \delta_{i2}) \right\}  \\
    + \log(\phi) + \log(\delta_{i1} \delta_{i2}) \Bigr],
\end{multline}
where $w(\Psi^{(k)};x_{i1},x_{i2}) = \E_{\hat{\Psi}^{(k)}}(W_i|X_1=x_{i1},X_2=x_{i2})$ denotes the expectation in \eqref{equ:EW_EM}, fixing the parameters $\Psi=\hat{\Psi}^{(k)}$. As a first step in maximizing $Q(\Psi ; \Psi^{(k)})$, one has that
\[
\frac{\partial}{\partial \phi} Q(\Psi ; \Psi^{(k)}) = \frac{n}{\phi} - \frac{1}{\phi (1-\phi)} \sum_{i=1}^n w(\Psi^{(k)};x_{i1},x_{i2}) =0,
\]
which leads to $\hat{\phi}^{(k+1)} = 1-\bar{w}(\Psi^{(k)};x_{i1},x_{i2})$, where $\bar{w}(\Psi^{(k)};x_{i1},x_{i2}) = \sum_{i=1}^n w(\Psi^{(k)};x_{i1},x_{i2}) /n$.

It is clear from \eqref{equ:Q_em} that $(\hat{\Lambda}^{(k+1)},\hat{\theta}^{(k+1)})$ is obtained by optimizing 
\begin{equation}\label{equ:M_em_step1}
\sum_{i=1}^n  w(\Psi^{(k)};x_{i1},x_{i2}) \log f_{\Lambda,\theta}(x_{i1},x_{i2}).
\end{equation}
As shown in \cite{Schulz/Genest/Mesfioui:2021}, one can decompose $\log f_{\Lambda,\theta}(x_{1},x_{2})$ into marginal components and a remaining dependence term viz.
\begin{equation}\label{equ:ll_decompose}
    \log f_{\Lambda,\theta}(x_{1},x_{2}) = \ell_1(\lambda_1;x_1) + \ell_2(\lambda_2;x_2) + \ell_D(\Lambda,\theta;x_1,x_2),
\end{equation}
where $\ell_j(\lambda_j;x_j)=\log g_{\lambda_j}(x_j)$ for $j=1,2$, and $\ell_D(\Lambda,\theta;x_1,x_2)$ is a remainder term stemming from the underlying dependence structure. The latter can be written as
\[
\ell_D(\Lambda,\theta;x_1,x_2)=\ln \left\{ \sum_{z_1=0}^{x_1} \sum_{z_2=0}^{x_2} \kappa_{\Lambda,\theta}(z_1,z_2;x_1,x_2) c_{\Lambda,\theta}(z_1,z_2) \right\}
\]
where $c_{\Lambda,\theta}$ is used to represent either $c_{\Lambda,\theta}^+$ or $c_{\Lambda,\theta}^-$ in the case of positive and negative dependence, respectively, and 
\[
\kappa_{\Lambda,\theta}(z_1,z_2;x_1,x_2) = \prod_{j=1}^2 {x_j \choose z_j} \theta^{z_j} (1-\theta)^{x_j-z_j} / g_{\theta\lambda_j}(z_j) .
\]
Substituting \eqref{equ:ll_decompose} into \eqref{equ:M_em_step1} leads to
\begin{equation}\label{equ:EM_ll_decompose}
    \sum_{i=1}^n  w(\Psi^{(k)};x_{i1},x_{i2})\left\{ \ell_1(\lambda_1;x_{i1}) + \ell_2(\lambda_2;x_{i2})  + \ell_D(\Lambda,\theta;x_{i1},x_{i2}) \right\}.
\end{equation}

While updates for the parameters $\Lambda^{(k)}$ and $\theta^{(k)}$ can then be obtained by directly optimizing \eqref{equ:EM_ll_decompose}, an alternative two-step procedure can be considered wherein the marginal parameters are first estimated based on the marginal contributions, and the dependence parameter is estimated thereafter. This two-step approach is commonly used to simplify the estimation procedure in multi-dimensional settings, and is particularly convenient in the context of copula models. For more details, see, e.g., \cite{Joe:2005}, who refers to the method as the inference function for the margins, or IFM, approach. More specifically, in the present setting, this amounts to solving 
\begin{equation}\label{EM:EE}
\begin{bmatrix}
    \frac{\partial}{\partial \lambda_1} \sum_{i=1}^n  w(\Psi^{(k)};x_{i1},x_{i2})  \ell_1(\lambda_1;x_{i1}) \\[1em]
    \frac{\partial}{\partial \lambda_2} \sum_{i=1}^n  w(\Psi^{(k)};x_{i1},x_{i2}) \ell_2(\lambda_2;x_{i2}) \\[1em]
    \frac{\partial}{\partial \theta} \sum_{i=1}^n  w(\Psi^{(k)};x_{i1},x_{i2}) \ell_D(\Lambda,\theta;x_{i1},x_{i2}) \\[0.5em]
\end{bmatrix} = \mathbf{0} .
\end{equation}
From this, one can update the marginal Poisson rates according to a simple weighted average:
\[
\lambda_j^{(k)}= \frac{\sum_{i=1}^n  w(\Psi^{(k)};x_{i1},x_{i2}) x_{ij}}{\sum_{i=1}^n  w(\Psi^{(k)};x_{i1},x_{i2})}, \quad j=1,2 .
\]
The dependence parameter $\theta$ can then be found numerically viz
\[
\theta^{(k)} = \argmax_{\theta \in (0,1)}  \sum_{i=1}^n  w(\Psi^{(k)};x_{i1},x_{i2}) \ell_D(\Lambda^{(k)},\theta;x_{i1},x_{i2}).
\]
Beginning with appropriate initial values $\Psi^{(0)}$, such as the MoM estimates $\check{\Psi}$, parameter values are iteratively updated according to the EM algorithm until a convergence criterion is achieved in either $\Psi^{(k+1)}-\Psi^{(k)}$, or in terms of the evaluated log-likelihood.

In the bivariate setting, maximum likelihood estimation involves a three-dimensional optimization, since $\phi$ has an explicit solution, and is thus numerically feasible. In such settings, it may not be necessary to consider the EM algorithm as outlined here. However, in higher dimensions, such as that presented in Section~\ref{sec: multivariateEX}, directly optimizing the likelihood will become increasingly complex and in such cases alternative approaches may be necessary. The EM algorithm detailed here breaks down the problem into a set of closed-form solutions for updating the parameters $\phi$ and $\Lambda$, and only a univariate optimization is required in solving for $\theta$.

\section{Simulations}\label{sec:simulations}

Comprehensive simulations were conducted in order to assess the performance of the estimation methods described in Section~\ref{sec:3.Estimation}. Several scenarios were considered with varying specifications, sepcifically, sample sizes $n \in \{100, 500, 1000\}$, and parameter values $\Lambda \in \{ (1,2), (5,10)\}$, $\phi \in \{0.1, 0.5, 0.9\}$, and $\theta \in \{0.25, 0.75\}$. Each setting was explored in both the models for positive and negative dependence, yielding a total of $72$ distinct scenarios, each of which were replicated over $R=500$ iterations.  Algorithm~\ref{alg:data-generating} was used to generate data for given parameter values $\Psi=(\Lambda,\theta,\phi)$ and sample size $n$. All simulations were carried out in $\mathsf{R}$. The \texttt{uniroot} function was used to find the method of moments estimator $\check{\theta}$. Maximum likelihood estimation was carried out using the \texttt{optim} function with the \texttt{Nelder-Mead} method and setting the maximum number of iterations to $1000$. While the method of moment estimates $\check{\Psi}$ were used as initial values in the $\mathcal{BZIP}^+$ model, the choice of appropriate starting points in the case of negative dependence was more challenging; this will be further discussed in Section~\ref{sec:simnegative}. Two implementations of maximum likelihood estimation were considered in the simulations, as described in Section~\ref{sec:ML}: The first is based on simultaneously optimizing the likelihood over all parameters $\Psi=(\Lambda,\theta,\phi)$, whereas the second considers a two-step approach wherein $(\hat{\Lambda},\hat{\theta})$ are first determined via \eqref{loglik2} and $\hat{\phi}$ is then given by $\eqref{equ:phiest}$. 

\begin{algorithm}
\begin{algorithmic}[1]
\caption{Data generation for each $i \in \{1,\ldots,n\}$}
\label{alg:data-generating}
    \State Generate $(U_i, V_{i1}, V_{i2}) \overset{iid}{\sim} \mathcal{U}(0,1)$
    \State Generate $W_i \sim \text{Bernoulli}(1-\phi)$
    \State Set $(Y_{i1},Y_{i2})=(G_{(1-\theta) \lambda_1}^{-1}(V_{i1}),G_{(1-\theta) \lambda_2}^{-1}(V_{i2}))$
    \State Set $(Z_{i1},Z_{i2})$
    
    \begin{itemize}
        \item[(a)] in the model for positive dependence, set $(Z_{i1},Z_{i2})=(G_{\theta \lambda_1}^{-1}(U_i),G_{\theta \lambda_2}^{-1}(U_i))$ 
        \item[(b)] in the model for negative dependence, set $(Z_{i1},Z_{i2})=(G_{\theta \lambda_1}^{-1}(U_i),G_{\theta \lambda_2}^{-1}(1-U_i))$
    \end{itemize}
    \State Set $(T_{i1}, T_{i2})=(Y_{i1}+Z_{i1},Y_{i2}+Z_{i2})$
    \State Set $(X_{i1},X_{i2})=W_i(T_{i1},T_{i2})$
\end{algorithmic}
\end{algorithm}

In each plot, estimates from the method of moments (MoM), two-step maximum likelihood implementation (MLE2) and one-step maximum likelihood implementation (MLE) are provided, with a solid horizontal line representing the true parameter value. In some instances, method of moments estimation was not feasible. As previously described, this happens whenever one, or both, components takes on only the values of $0$ or $1$. Such replications were omitted from the figures across all estimation methods depicted to allow for a more meaningful comparison. As such, each figure provides the number of replications $R$ shown in the corresponding plot. Further details will be given in the subsequent sections.

\subsection{Model for positive dependence}\label{sec:simpositive}

\begin{figure}[H]
    \centering
  \includegraphics[width=0.7\linewidth]{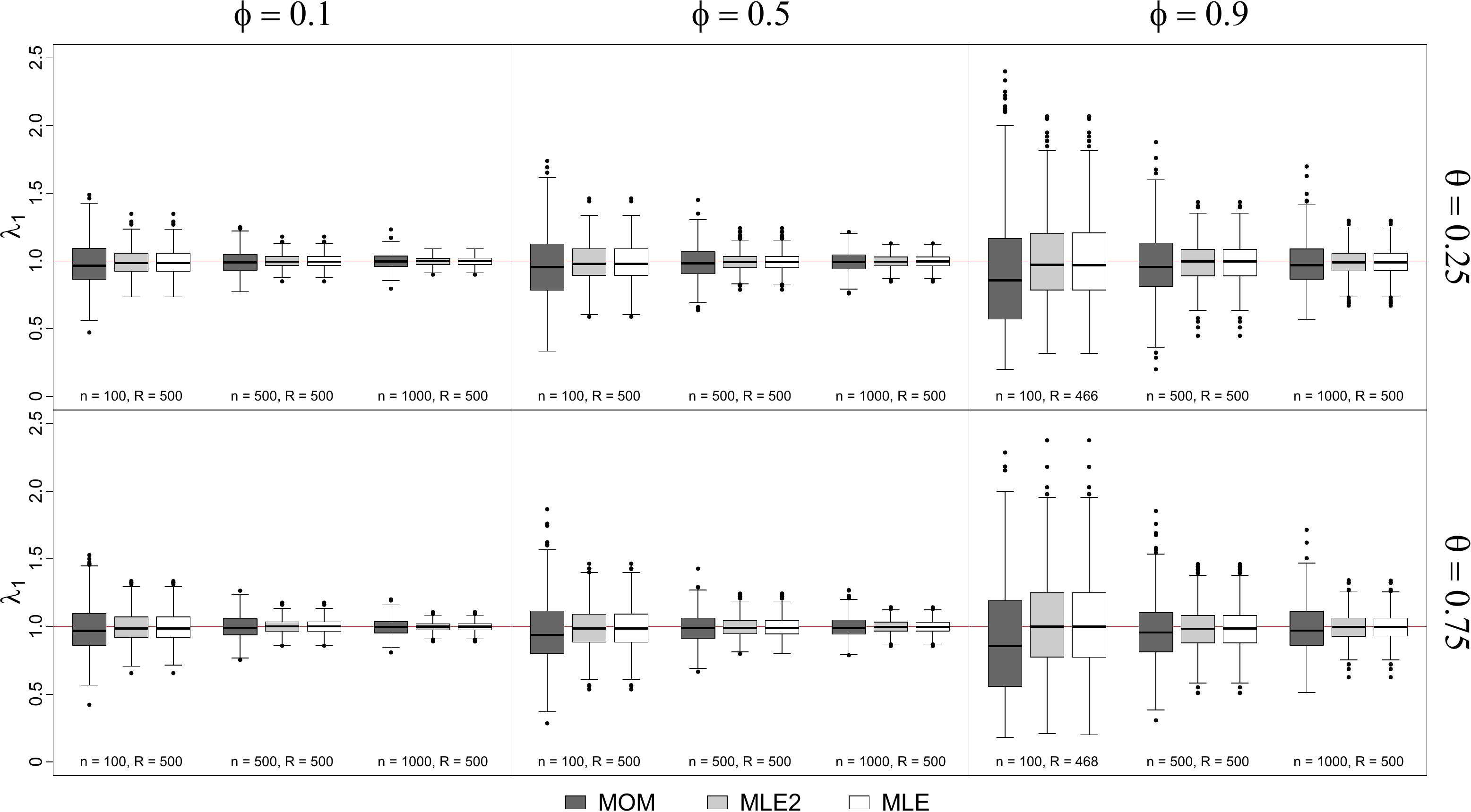}

\caption{Estimation of $\lambda_1$ in the $\mathcal{BZIP}^+$ model across scenarios with \(\Lambda = (1,2)\). The top and bottom rows respectively display the results for weaker levels of dependence (\(\theta = 0.25\)), and stronger levels of dependence $(\theta=0.75)$, while the columns display increasing zero-inflation rates (from left to right, $\phi=0.1$, $\phi=0.5$ and $\phi=0.9$). Each plot shows the results for $n \in \{100, 500, 1000\}$, across $R$ replications, with the true parameter value indicated by the red horizontal line.}
 \label{fig:pos_lambda1_1_2}
\end{figure}

\begin{figure}[H]
\centering
    \includegraphics[width=0.7\linewidth]{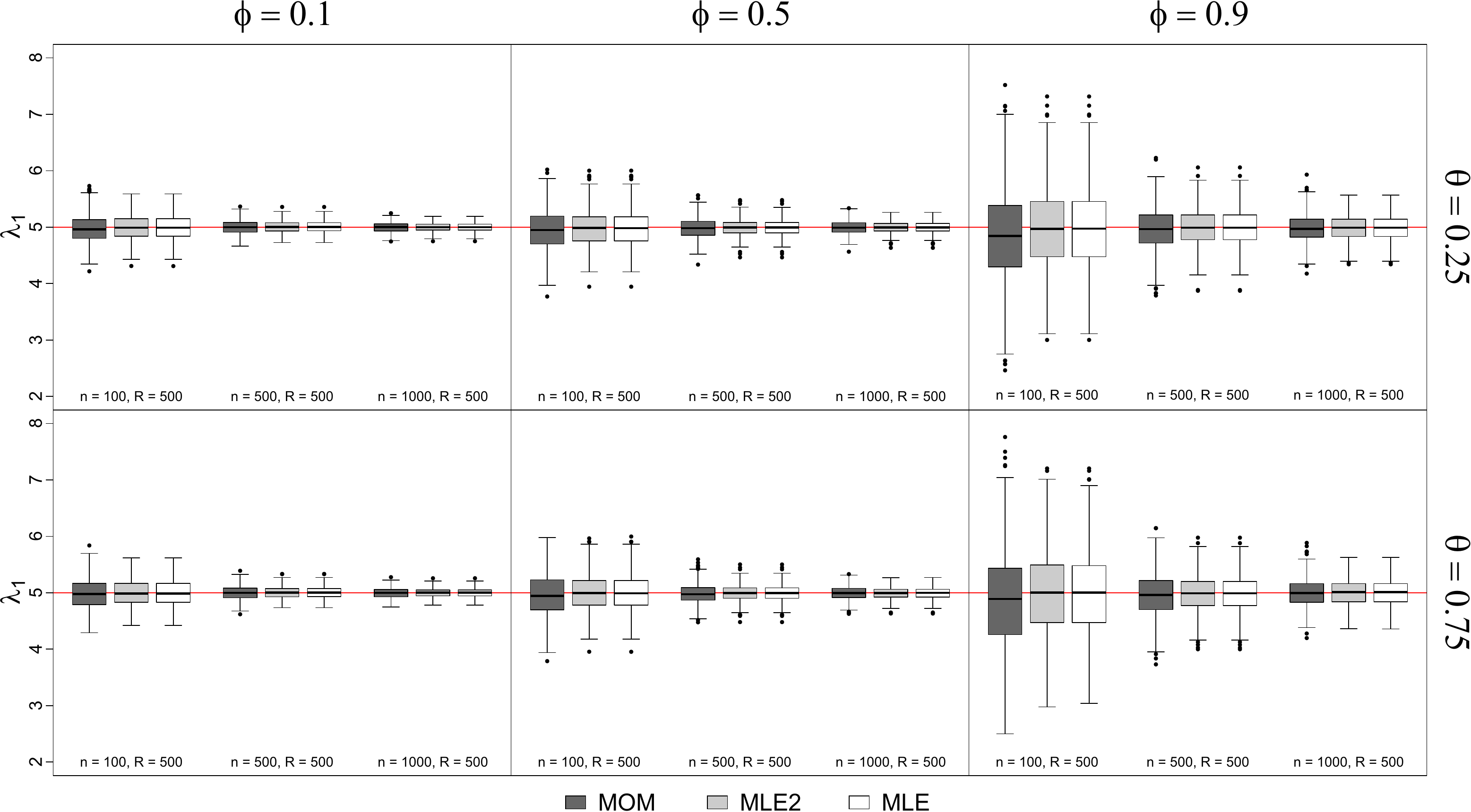}

\caption{Estimation of $\lambda_1$ in the $\mathcal{BZIP}^+$ model across scenarios with \(\Lambda = (5,10)\). The top and bottom rows respectively display the results for weaker levels of dependence (\(\theta = 0.25\)), and stronger levels of dependence $(\theta=0.75)$, while the columns display increasing zero-inflation rates (from left to right, $\phi=0.1$, $\phi=0.5$ and $\phi=0.9$). Each plot shows the results for $n \in \{100, 500, 1000\}$, across $R$ replications, with the true parameter value indicated by the red horizontal line.}
 \label{fig:pos_lambda1_5_10}
\end{figure}

\begin{figure}[H]
\centering
\includegraphics[width=0.7\linewidth]{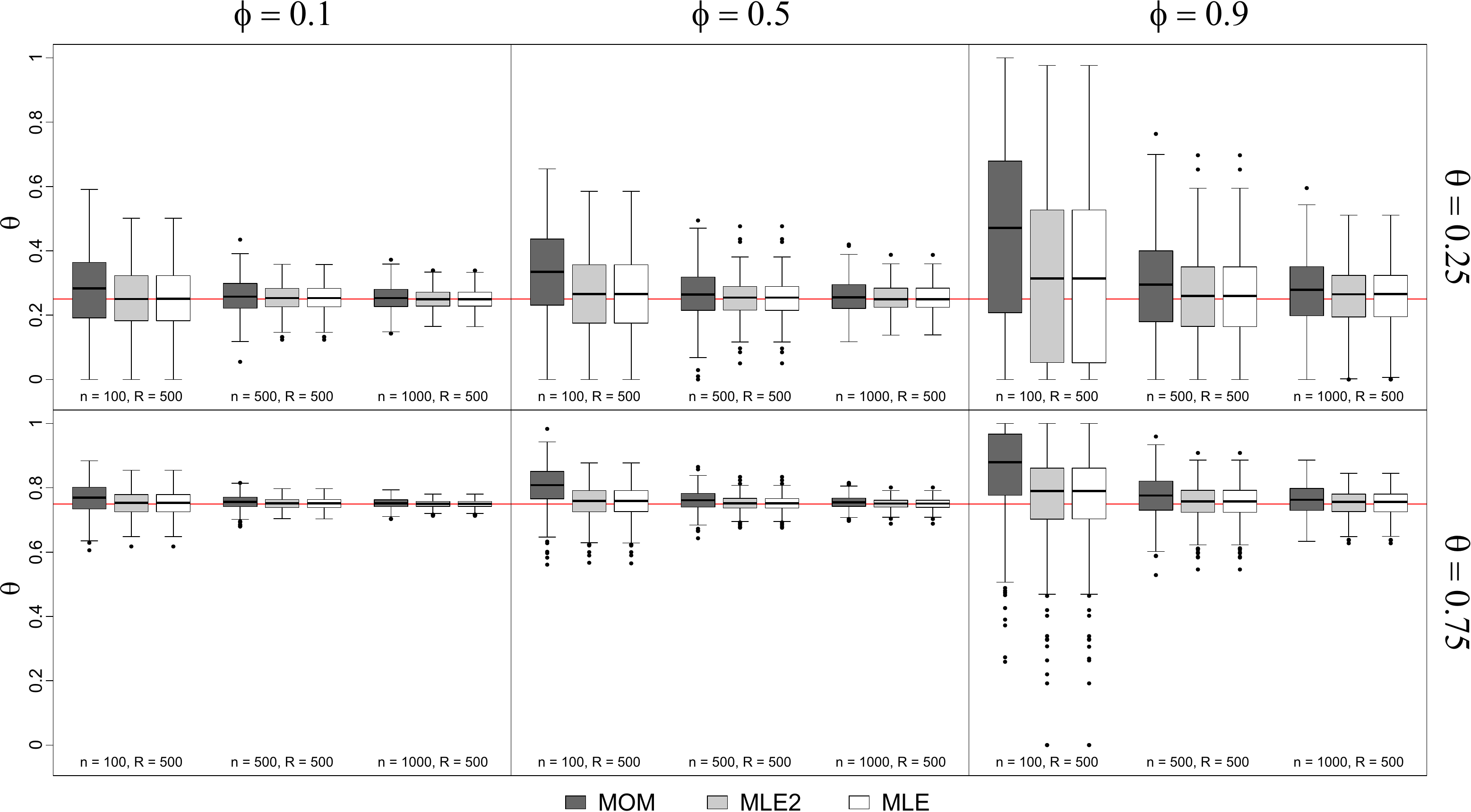}
 
\caption{Estimation of $\theta$ in the $\mathcal{BZIP}^+$ model across scenarios where $\Lambda=(5,10)$. The top and bottom rows respectively display the results for weaker levels of dependence (\(\theta = 0.25\)), and stronger levels of dependence $(\theta=0.75)$, while the columns display increasing zero-inflation rates (from left to right, $\phi=0.1$, $\phi=0.5$ and $\phi=0.9$). Each plot shows the results for $n \in \{100, 500, 1000\}$, across $R$ replications, with the true parameter value indicated by the red horizontal line.}
 \label{fig:pos_theta_5_10}
\end{figure}

A representative subset of the results across the 36 scenarios in the $\mathcal{BZIP}^+$ model are presented in Figures~\ref{fig:pos_lambda1_1_2} trough~\ref{fig:pos_theta_5_10}. In particular, Figure~\ref{fig:pos_lambda1_1_2} shows the estimated values of $\lambda_1$ in scenarios with $\Lambda=(1,2)$, while results for scenarios with $\Lambda=(5,10)$ are shown in Figure~\ref{fig:pos_lambda1_5_10}. Overall, the three approaches (MoM, MLE, MLE2) perform well for estimating the marginal Poisson rate, with improved results as $n$ increases. There is slightly more variability in the MoM estimates, while the two ML methods are indistinguishable. The results also suggest stability across the dependence levels, namely as $\theta$ varies from $0.25$ (shown in the top row) to $0.75$ (shown in the bottom row). On the other hand, there is seemingly more variability in estimating $\lambda_1$ as the inflation rate $\phi$ increases, as seen across the columns of Figure~\ref{fig:pos_lambda1_1_2} when $\phi$ takes on values $\{0.1,0.5, 0.9\}$. Figure~\ref{fig:pos_lambda1_5_10} shows a similar pattern in the case with larger marginal Poisson rates. This is not surprising as a larger $\phi$ leads to a greater proportion of $(0,0)$ in the generated samples. Indeed, as shown in equation~\ref{loglik2}, all of the information pertaining to the marginal Poisson rates is in the subset $\{i \notin  M_0 \}$. Note that while not shown, estimation of $\lambda_2$ led to similar patterns as well.

\begin{figure}[H]
    \centering
     \includegraphics[width=0.5\textwidth]{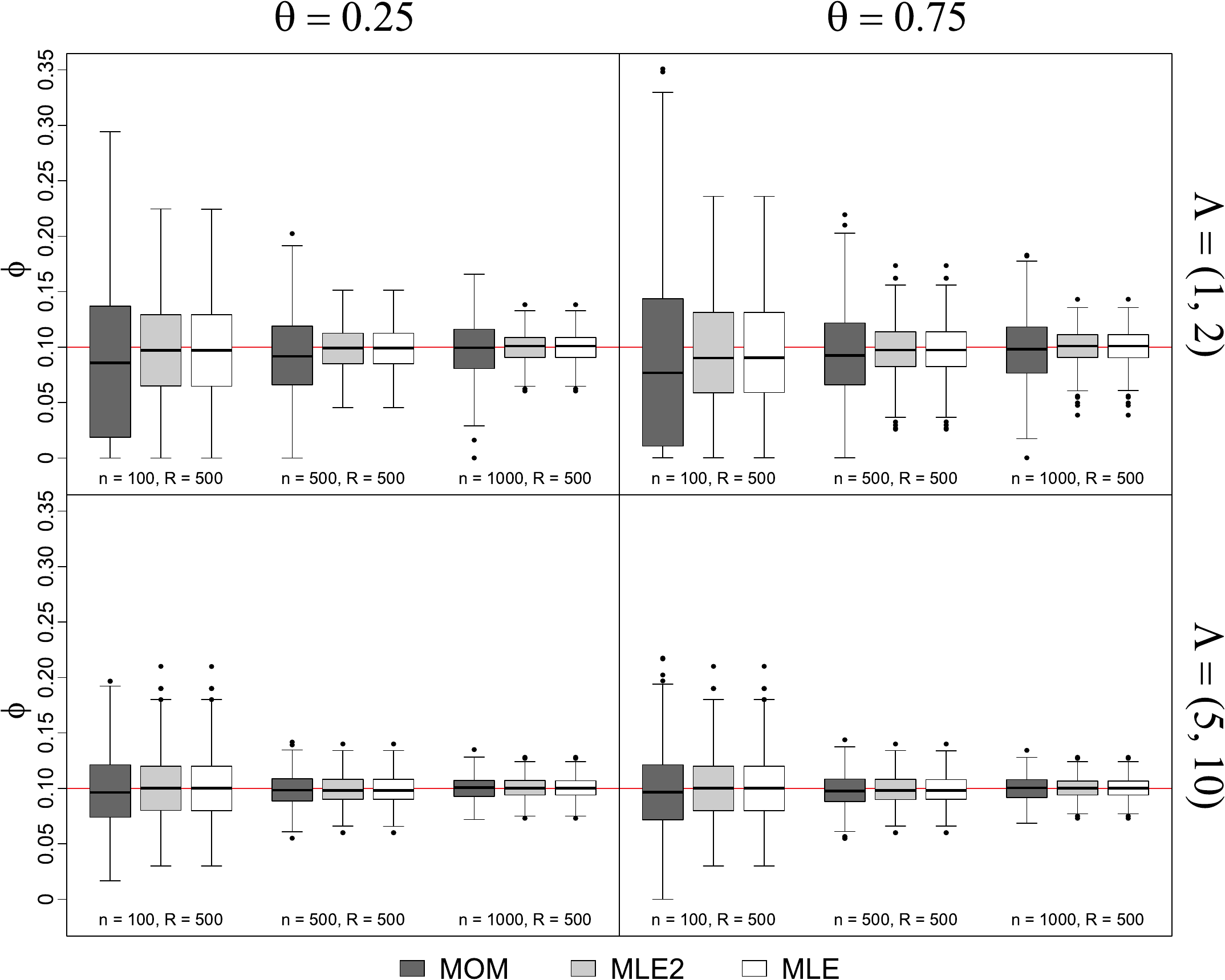}

    \caption{ Estimation of $\phi$ in the $\mathcal{BZIP}^+$ model across scenarios where $\phi=0.1$. The top and bottom rows respectively display the results for smaller Poisson rates of \(\Lambda = (1,2)\), and large Poisson rates of \(\Lambda = (5,10)\). The left panel provide the results for weaker levels of dependence ($\theta=0.25$), while the right panel shows that for stronger levels of dependence ($\theta=0.75$). Each plot shows the results for $n \in \{100, 500, 1000\}$, across $R$ replications, with the true parameter value indicated by the red horizontal line.
    }
    \label{fig:pos_phi_0.1}
\end{figure}

\begin{figure}[H]
    \centering
    \includegraphics[width=0.5\textwidth]{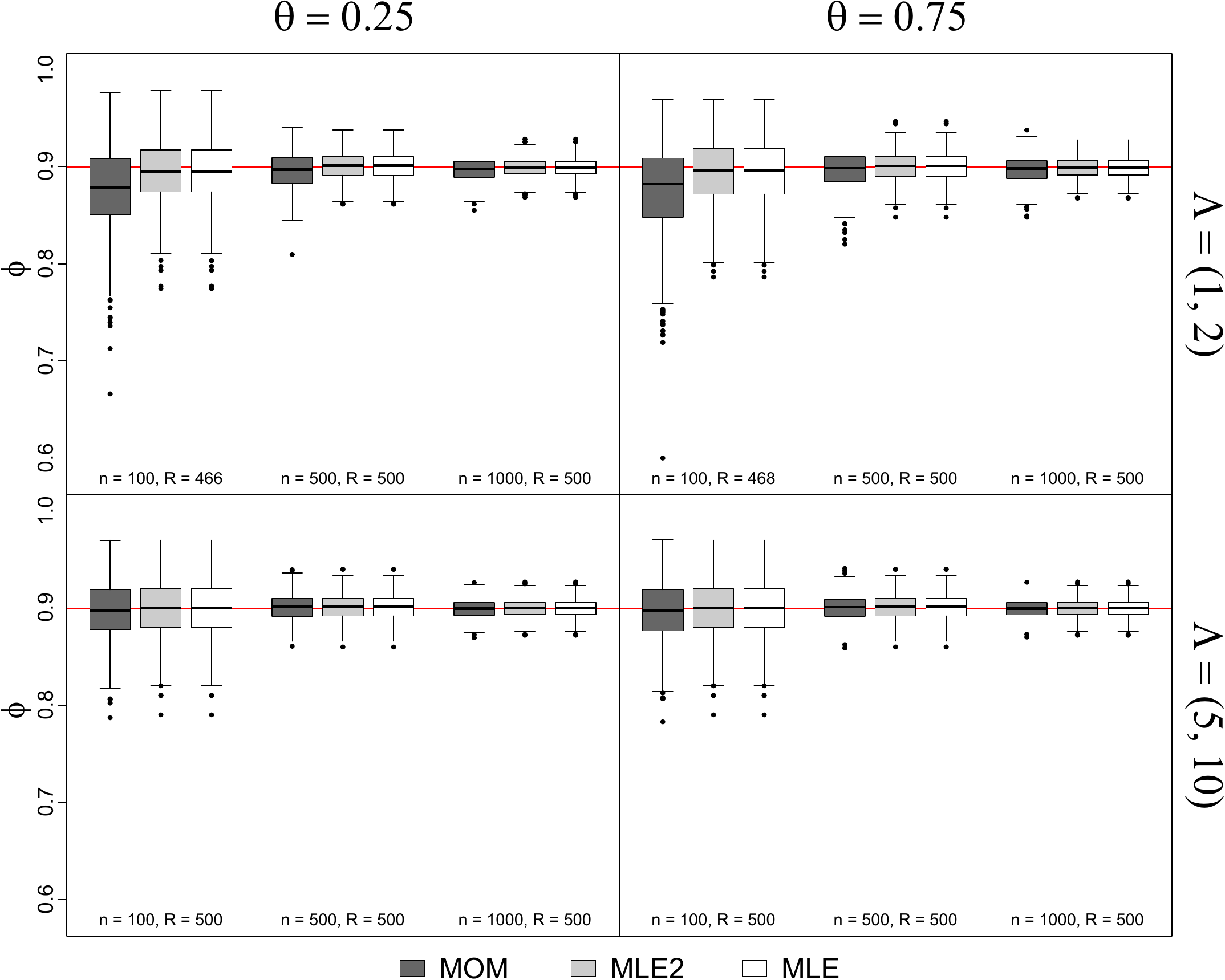}
    
    \caption{Estimation of $\phi$ in the $\mathcal{BZIP}^+$ model across scenarios where $\phi=0.9$. The top and bottom rows respectively display the results for smaller Poisson rates (\(\Lambda = (1,2)\)) and large Poisson rates (\(\Lambda = (5,10)\)). The left panel provide the results for weaker levels of dependence ($\theta=0.25$), while the right panel shows that for stronger levels of dependence ($\theta=0.75$). Each plot shows the results for $n \in \{100, 500, 1000\}$, across $R$ replications, with the true parameter value indicated by the red horizontal line.}
    \label{fig:pos_phi_0.9}
\end{figure}

Figures~\ref{fig:pos_phi_0.1} and~\ref{fig:pos_phi_0.9} provide the estimation results for $\phi$ in the scenarios with $\phi=0.1$ and $\phi=0.9$, respectively. The plots seem to suggest a slight improvement in estimating $\phi$ when the marginal rates are larger (going from $\Lambda=(1,2)$ in the top row to $\Lambda=(5,10)$ in the bottom row). This phenomenon was more pronounced for $\phi=0.1$, and for smaller $n$. Note that when the marginal Poisson rates are smaller, the probability of double-zeros stemming from the latent count variables, i.e. $\Prob(T_1=0,T_2=0)$, increases, thereby increasing $m_0$. Across the scenarios summarized in Figures~\ref{fig:pos_phi_0.1} and~\ref{fig:pos_phi_0.9}, it seems that estimation of $\phi$ is unaffected by the strength of dependence, as measured through $\theta$. Note that some iterations yielded negative results for the MoM and MLE2 estimates of $\phi$. This happened a total of $1.2\%$ times for MoM, and $0.1\%$ times for MLE2 over the total of $18000$ replications considered in the $\mathcal{BZIP}^+$, and mostly occurred in small samples, where both $\phi$ and $\Lambda$ were smaller.

\begin{figure}[H]
    \centering
    \includegraphics[width=0.7\linewidth]{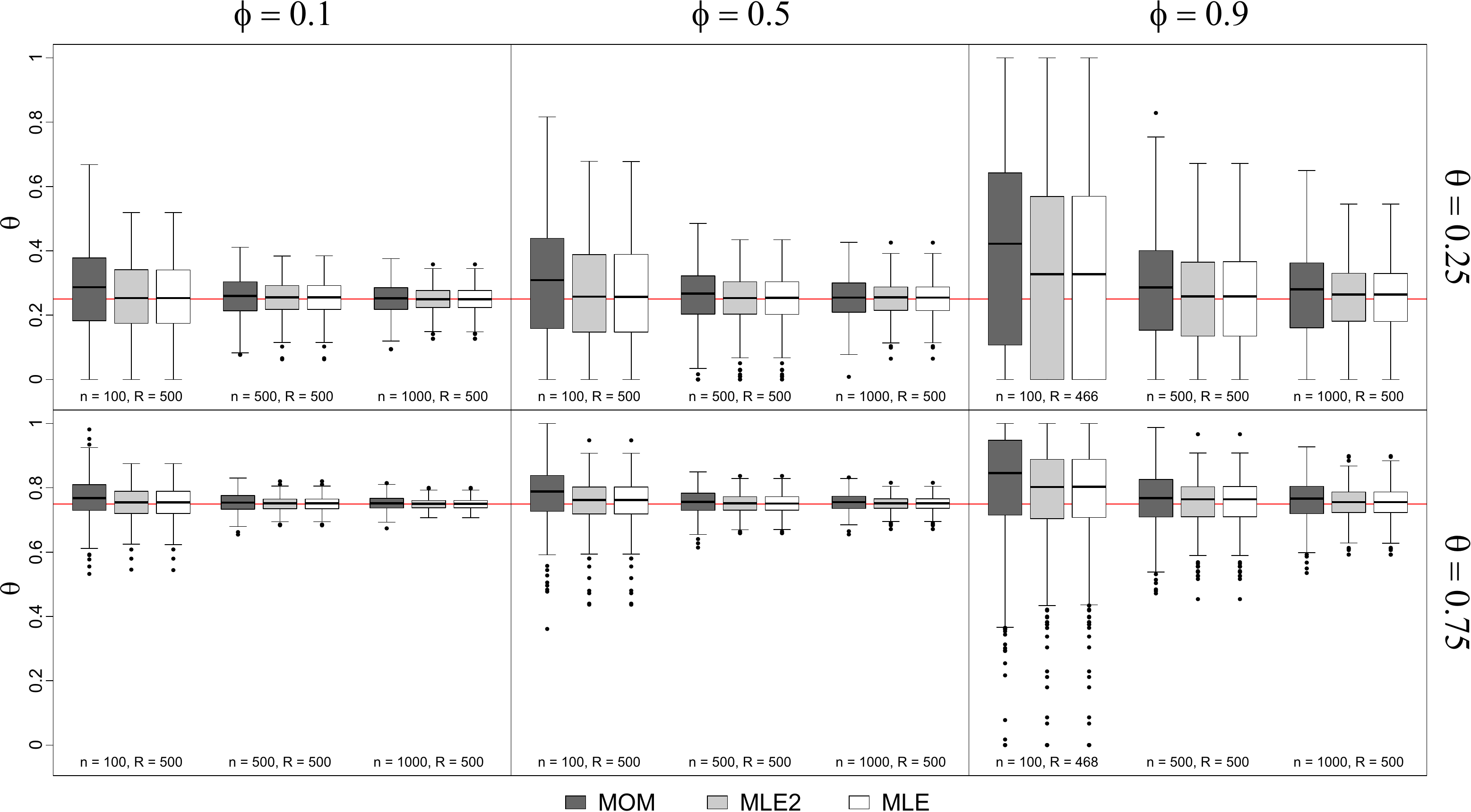}

\caption{Estimation of $\theta$ in the $\mathcal{BZIP}^+$ model across scenarios where $\Lambda=(1,2)$. The top and bottom rows respectively display the results for weaker levels of dependence (\(\theta = 0.25\)), and stronger levels of dependence $(\theta=0.75)$, while the columns display increasing zero-inflation rates (from left to right, $\phi=0.1$, $\phi=0.5$ and $\phi=0.9$). Each plot shows the results for $n \in \{100, 500, 1000\}$, across $R$ replications, with the true parameter value indicated by the red horizontal line.}
 \label{fig:pos_theta_1_2}
\end{figure}

Figures~\ref{fig:pos_theta_1_2} and~\ref{fig:pos_theta_5_10} display the resulting estimates for $\theta$ in settings where $\Lambda=(1,2)$ and $\Lambda=(5,10)$, respectively. The results show more volatility in estimating $\theta$ as $\phi$ increases, as seen going from left to right in each respective row in the plots. This is similar to what was seen in estimating $\lambda_1$: increasing $\phi$ results in less information for estimating the latent Poisson count parameters $(\Lambda,\theta)$. Again, this is in line with equation~\eqref{loglik2}, where it is clear that a bigger sample size $n$ may be required to have sufficient information to estimate $(\Lambda,\theta)$ when $\phi$ is large and hence $m_0$ is large. There is slightly less variability in the estimates for $\theta$ in settings with strong levels of dependence (i.e., in scenarios with $\theta=0.75$ in comparison to $\theta=0.25$). The figures further suggest that estimation of $\theta$ is not impacted by the marginal Poisson rates. 

The results for the $\mathcal{BZIP^+}$ model show that the two maximum likelihood approaches yield essentially identical results throughout the simulations. In terms of root mean squared error (RMSE), the likelihood-based methods always outperformed the method of moments across all scenarios. 

As previously mentioned, numerical challenges can sometimes arise in implementing the estimation procedures. In settings where the Poisson marginal rates are small (here, $\Lambda=(1,2)$) and the inflation factor is large (here $\phi=0.9$), it sometimes occurred that the generated sample had at least one margin consisting solely of values $0$ or $1$, rendering estimating impossible via the MoM and challenging for MLE. In practice, in such settings, a Bernoulli model would likely be considered for the margin, rather than a zero-inflated Poisson. In the simulation study considered here, this occurred for merely $0.4\%$ of the $18000$ iterations, and only when the sample size was small ($n=100$). Note that these replications were omitted from the plots, as indicated by an $R$ less than $500$. Further problems may occur in method of moments estimation wherein the implied solution to $S_{12} = s^+_{\check{\Lambda},\check{\phi}}(\theta)$ yields a value of $\theta$ extending beyond the $(0,1)$ interval. In such cases, the estimate was capped at either $0$ or $1$, as the case may be. In the settings considered here, $2.0\%$ and $1.2\%$ of the $18000$ iterations were respectively capped at $0$ and $1$. Not surprisingly, such instances tended to occur in smaller sample sizes, and for larger $\phi$. 

\begin{figure}[H]
    \centering
    \includegraphics[width=0.5\textwidth]{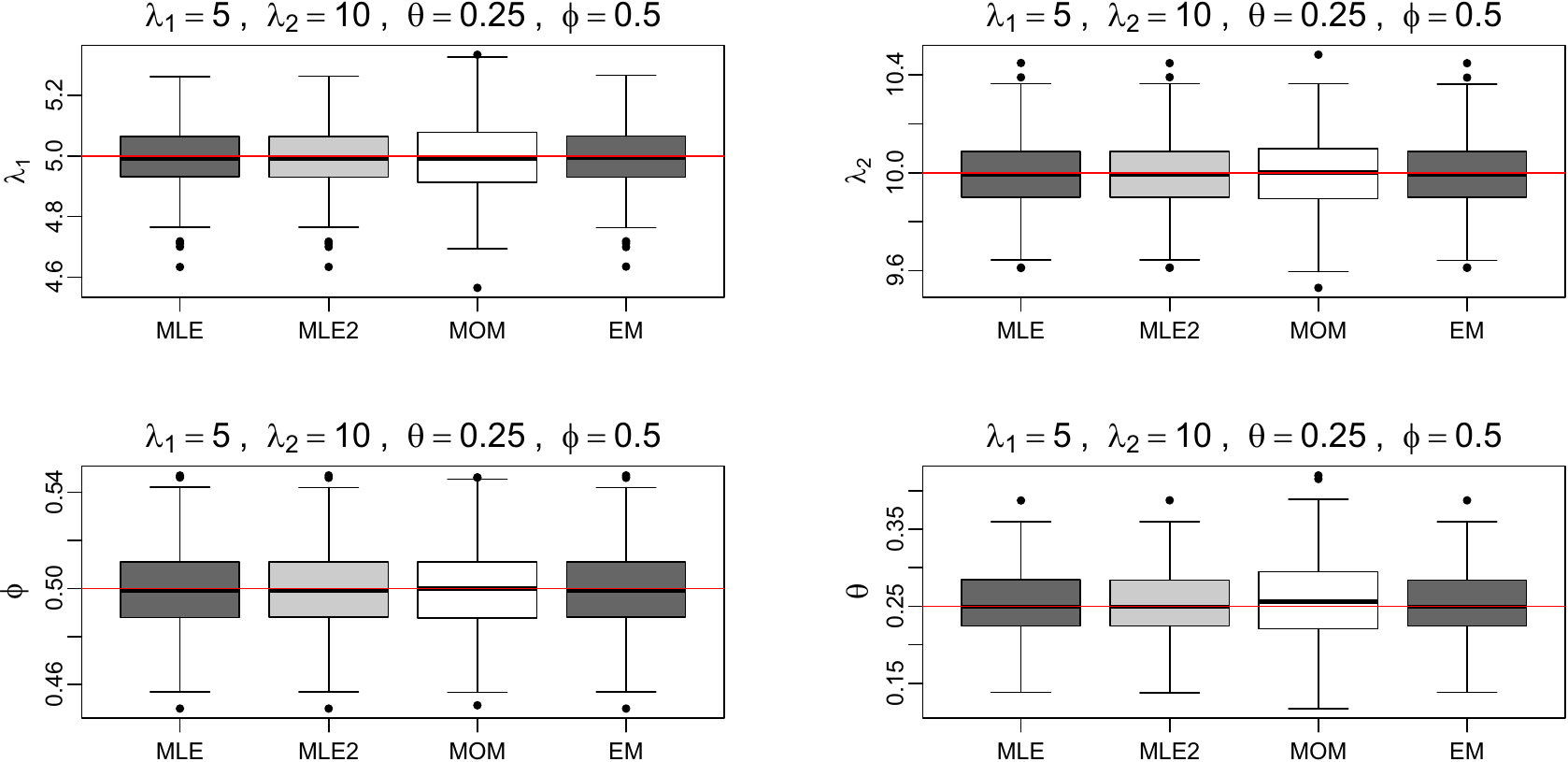}
    
    \caption{Estimation results for the EM algorithm in the $\mathcal{BZIP}^+$ model: results for $\lambda_1$ are shown at the top left, $\lambda_2$ at the top right, $\phi$ at the bottom left, and $\theta$ at the bottom right. The estimates provided are based on a sample of $n=1000$ observations, across $R=500$ replications. The true values, $\Lambda = (5,10)$, $\theta = 0.25$, and $\phi = 0.5$, are shown in red in each respective plot.    
    }
    \label{fig:EM}
\end{figure}

Finally, performance of the EM algorithm is illustrated in Figure~\ref{fig:EM} for the specific scenario with $\Lambda=(5,10)$, $\phi=0.5$ and $\theta=0.25$. These results are based on using the MoM estimates as a starting value. Overall, the algorithm is shown to yield good results, but took longer to run in comparison to direct ML estimation.

\subsection{Model for negative dependence}\label{sec:simnegative}

\begin{figure}[H]
\centering
\includegraphics[width=0.7\linewidth]{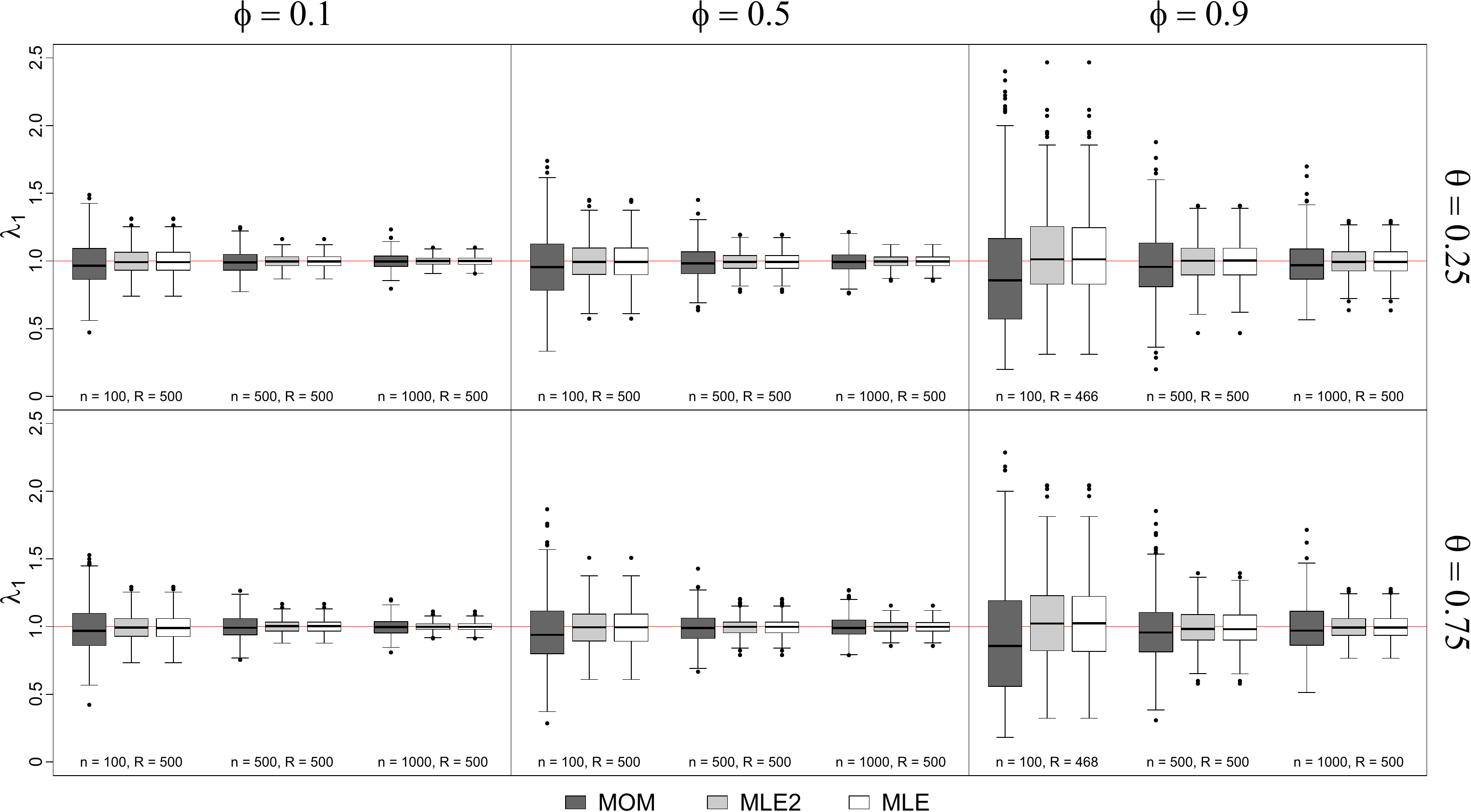}

\caption{Estimation of $\lambda_1$ in the $\mathcal{BZIP}^-$ model across scenarios with \(\Lambda = (1,2)\). The top and bottom rows respectively display the results for weaker levels of dependence (\(\theta = 0.25\)), and stronger levels of dependence $(\theta=0.75)$, while the columns display increasing zero-inflation rates (from left to right, $\phi=0.1$, $\phi=0.5$ and $\phi=0.9$). Each plot shows the results for $n \in \{100, 500, 1000\}$, across $R$ replications, with the true parameter value indicated by the red horizontal line.}
 \label{fig:neg_lambda1_1_2}
\end{figure}

\begin{figure}[H]
\centering
\includegraphics[width=0.7\linewidth]{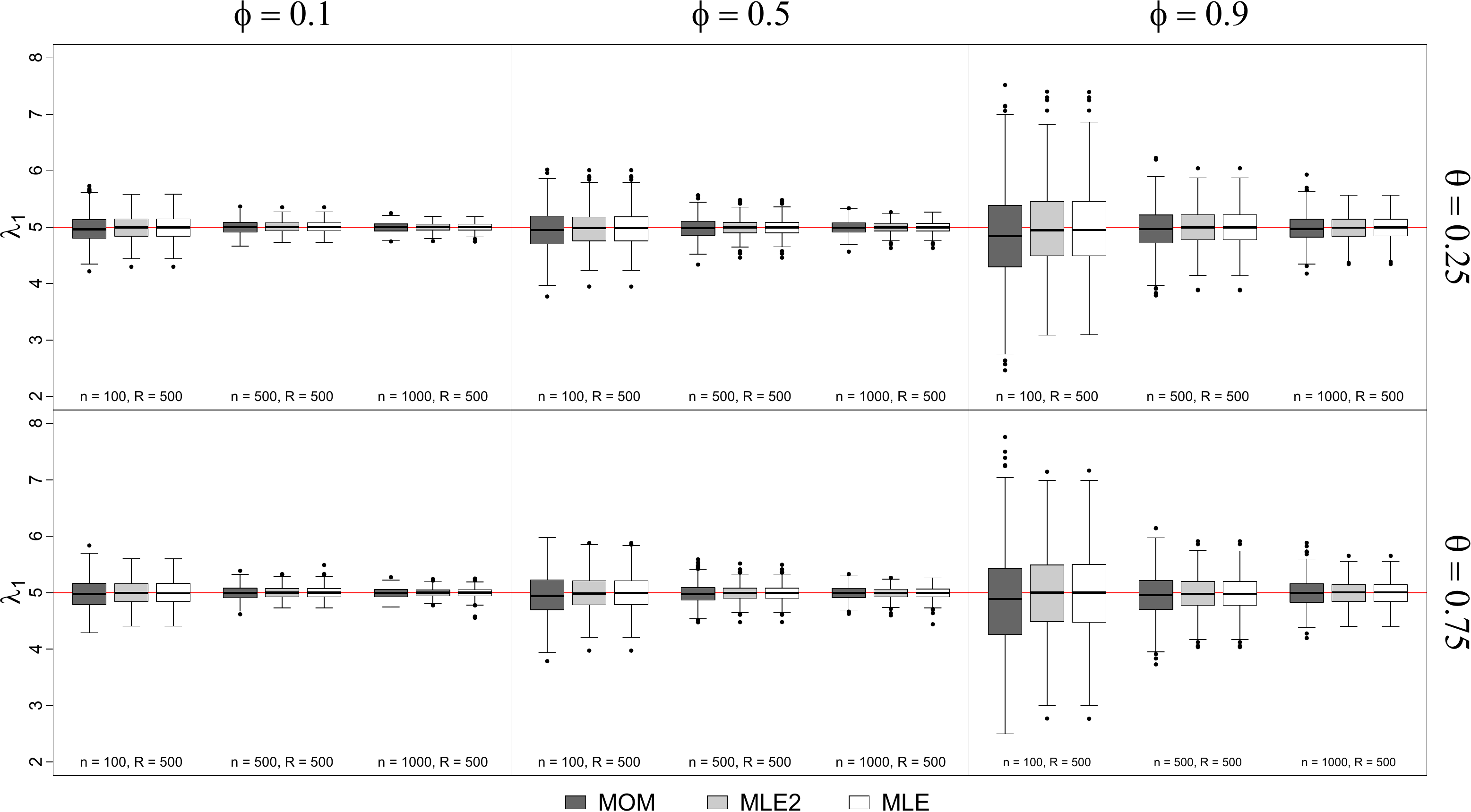}

\caption{Estimation of $\lambda_1$ in the $\mathcal{BZIP}^-$ model across scenarios with \(\Lambda = (5,10)\). The top and bottom rows respectively display the results for weaker levels of dependence (\(\theta = 0.25\)), and stronger levels of dependence $(\theta=0.75)$, while the columns display increasing zero-inflation rates (from left to right, $\phi=0.1$, $\phi=0.5$ and $\phi=0.9$). Each plot shows the results for $n \in \{100, 500, 1000\}$, across $R$ replications, with the true parameter value indicated by the red horizontal line.}
 \label{fig:neg_lambda1_5_10}
\end{figure}

\begin{figure}[H]
    \centering
   
        \centering
        \includegraphics[width=0.6\textwidth]{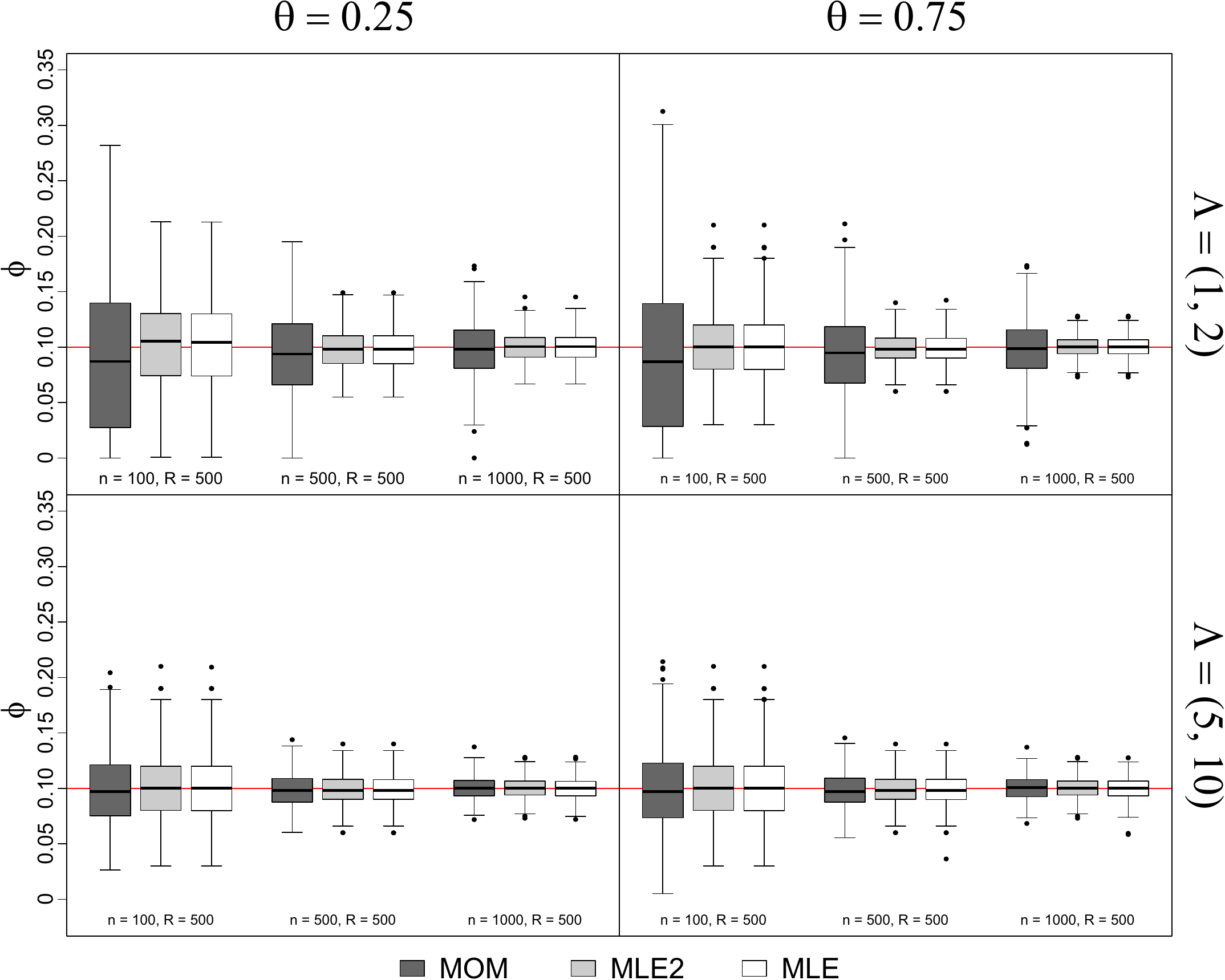}
    
    \caption{Estimation of $\phi$ in the $\mathcal{BZIP}^-$ model across scenarios where $\phi=0.1$. The top and bottom rows respectively display the results for smaller Poisson rates (\(\Lambda = (1,2)\)) and large Poisson rates (\(\Lambda = (5,10)\)). The left panel provide the results for weaker levels of dependence ($\theta=0.25$), while the right panel shows that for stronger levels of dependence ($\theta=0.75$). Each plot shows the results for $n \in \{100, 500, 1000\}$, across $R$ replications, with the true parameter value indicated by the red horizontal line. }
    \label{fig:neg_phi_0.1}
\end{figure}

\begin{figure}[H]
    \centering
   \includegraphics[width=0.6\textwidth]{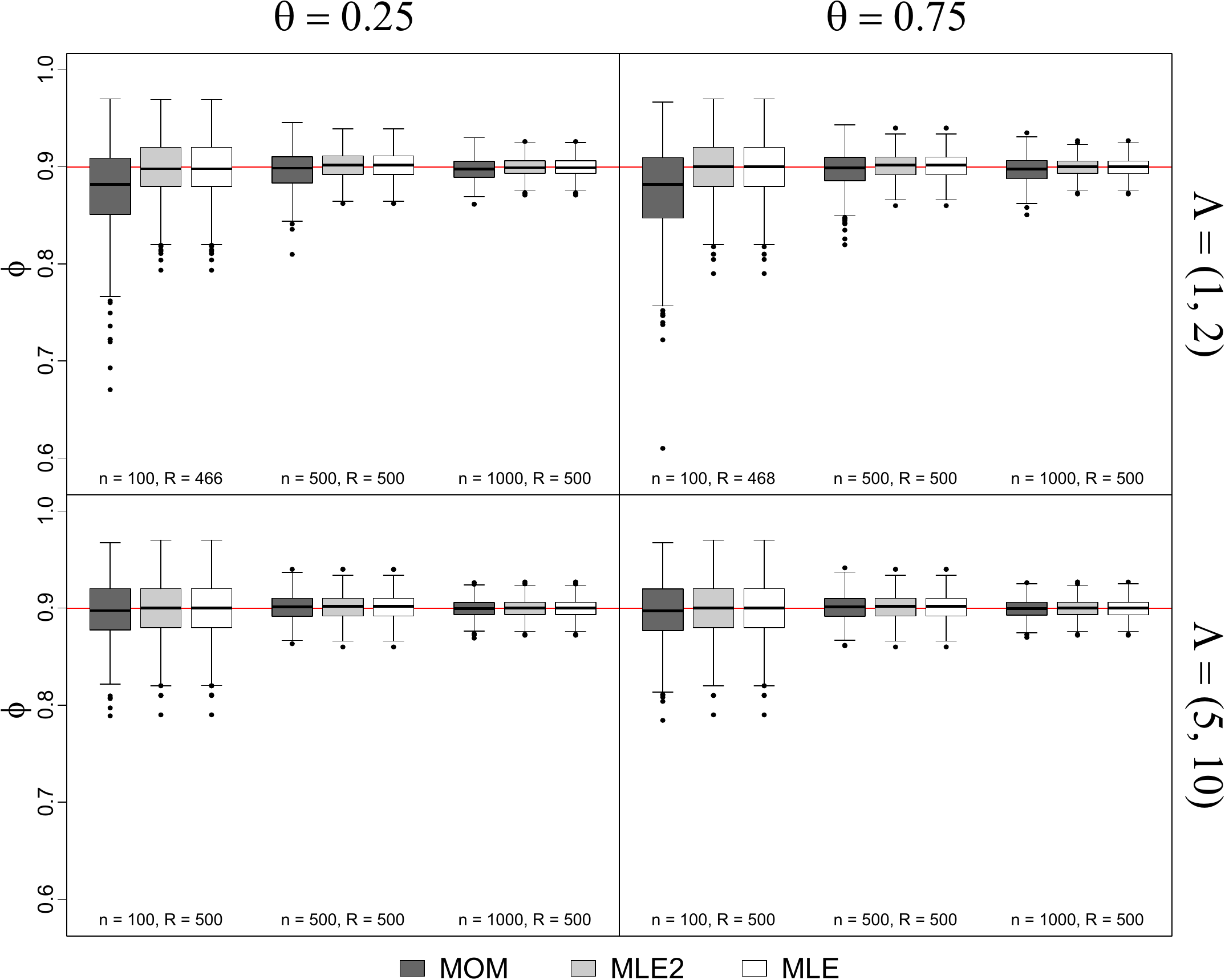}

    \caption{ Estimation of $\phi$ in the $\mathcal{BZIP}^-$ model across scenarios where $\phi=0.9$. The top and bottom rows respectively display the results for smaller Poisson rates (\(\Lambda = (1,2)\)) and large Poisson rates (\(\Lambda = (5,10)\)). The left panel provide the results for weaker levels of dependence ($\theta=0.25$), while the right panel shows that for stronger levels of dependence ($\theta=0.75$). Each plot shows the results for $n \in \{100, 500, 1000\}$, across $R$ replications, with the true parameter value indicated by the red horizontal line. }
    \label{fig:neg_phi_0.9}
\end{figure}

\begin{figure}[H]
\centering
  \includegraphics[width=0.7\linewidth]{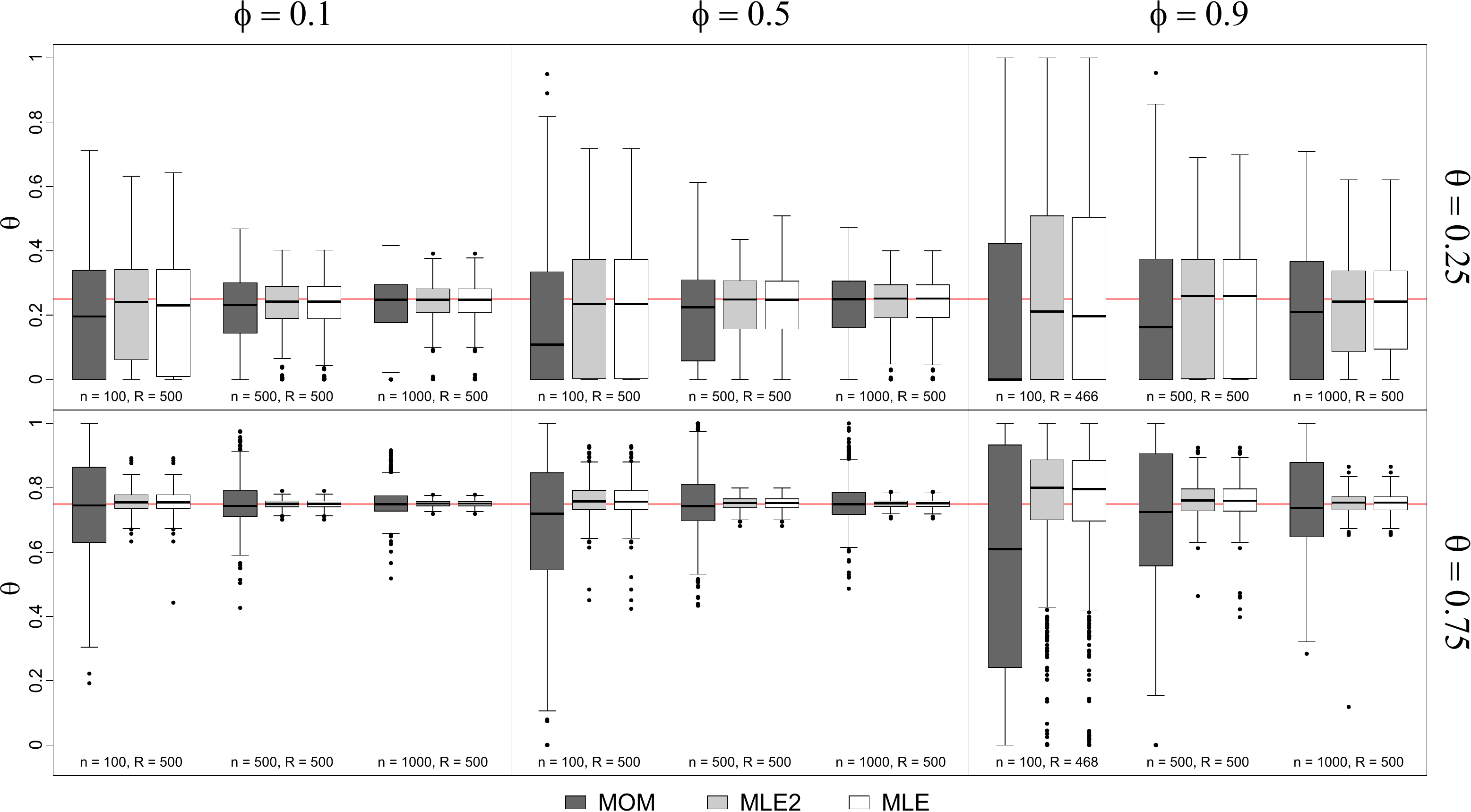}

\caption{Estimation of $\theta$ in the $\mathcal{BZIP}^-$ model across scenarios where $\Lambda=(1,2)$. The top and bottom rows respectively display the results for weaker levels of dependence (\(\theta = 0.25\)), and stronger levels of dependence $(\theta=0.75)$, while the columns display increasing zero-inflation rates (from left to right, $\phi=0.1$, $\phi=0.5$ and $\phi=0.9$). Each plot shows the results for $n \in \{100, 500, 1000\}$, across $R$ replications, with the true parameter value indicated by the red horizontal line. }
 \label{fig:neg_theta_1_2}
\end{figure}

\begin{figure}[H]
\centering
  \includegraphics[width=0.5\linewidth]{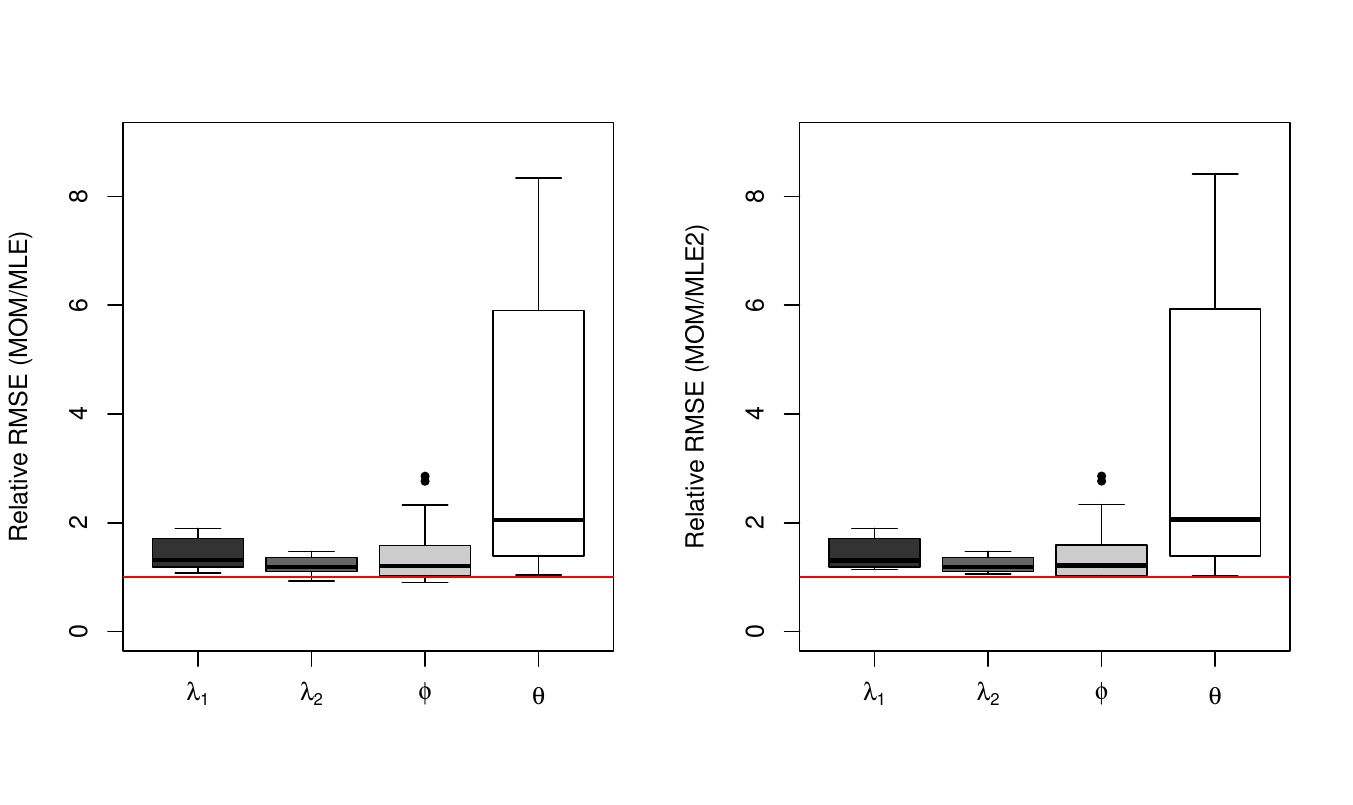}
\caption{The relative RMSE of the method of moment estimator over the maximum likelihood approach 1 (left) and 2 (right) for each parameter in the $\mathcal{BZIP}^-$ model}
 \label{fig:nd:RMSE_ND}
\end{figure}

The same subset of the $36$ distinct scenarios previously explored in the $\mathcal{BZIP^+}$ model are shown in the model for negative dependence in Figures~\ref{fig:neg_lambda1_1_2} through \ref{fig:neg_theta_5,10}. Overall, similar patterns arise in the $\mathcal{BZIP}^-$ model. In terms of the marginal Poisson rates, shown in Figures~\ref{fig:neg_lambda1_1_2} for $\Lambda=(1,2)$ and \ref{fig:neg_lambda1_5_10} for $\Lambda=(5,10)$, estimation of $\lambda_1$ is shown to be stable across $\theta$ levels, with increasing variability as $\phi$ increases. Analogously, estimation of $\theta$ is seemingly unaffected by $\Lambda$, but shows more volatility for increased values of $\phi$, as depicted in Figures~\ref{fig:neg_theta_1_2} for $\Lambda=(1,2)$ and \ref{fig:neg_theta_5,10} for $\Lambda=(5,10)$. As previously discussed, this stems from the fact that the information regarding the Poisson parameters $\Lambda$ and $\theta$ are entirely in the subset of $\{i \notin M_0\}$, as per equation \eqref{loglik2}. Estimation of the inflation rate $\phi$ also follows similar trends as those identified in the model for positive dependence, as shown in Figures~\ref{fig:neg_phi_0.1} for $\phi=0.1$ and \ref{fig:neg_phi_0.9} for $\phi=0.9$. In particular, estimation of $\phi$ was rather unaffected by the strength of dependence, as measured through $\theta$, and seemingly led to very slightly less volatility when $\Lambda$ is larger. As seen in the results for the $\mathcal{BZIP}^+$ model, in the setting with negative dependence, both likelihood-based methods yield essentially identical results. Moreover, at least one of the two ML methods always provided more accurate results in terms of RMSE in comparison to the method of moments; this is shown in Figure \ref{fig:nd:RMSE_ND}, which illustrates the relative RMSE for both the MLE and MLE2 methods with respect to the MoM.

While the results are similar in both the positive and negative dependence settings, estimation was more challenging in the $\mathcal{BZIP}^-$ model. Across the $18000$ iterations, $0.4\%$ of the generated samples resulted in at least one margin consisting entirely of the values $0$ or $1$, rendering MoM estimation impossible. As before, these replications are removed in all plots. Similar to the case of positivie dependence, method of moments estimation sometimes led to $\check{\phi}<0$; this occurred in $1.5\%$ of the total $18000$ iterations, and happened mainly in small samples when both $\phi$ and $\Lambda$ were small. Recall that an additional numerical issue in method of moments estimation arises when the sample covariance extends beyond the permissible range of $[s_{\check{\Lambda},\check{\phi}}^-(1),s_{\check{\Lambda},\check{\phi}}^-(0)]$, in which case $\check{\theta}$ is capped at $1$ or $0$, as the case may be. This phenomena happened more frequently in the model with negative dependence, with $13.3\%$ of the total of $18000$ replications resulting in $\check{\theta}=0$, and $4.5\%$ of the total iterations yielding $\check{\theta}=1$. Such issues were detected again for smaller sample sizes. 

Initializing the optimization procedures required for maximum likelihood estimation was more challenging in the $\mathcal{BZIP}^-$ model. Whereas in the positive dependence setting the MoM estimates were always used as starting values for $\theta^{(0)}$, this sometimes led to numerical issues in the model for negative dependence, occurring in $17.7\%$ of the total $18000$ iterations. As such, various initializing values were explored for the $\mathcal{BZIP}^-$ model, including $\theta^{(0)} \in \{0.1,\check{\theta},\theta_0,0.99\}$, where $\theta_0$ denotes the true parameter value. Among these choices, initializing $\theta^{(0)}=0.1$ was found to be numerically appropriate across all of the $18000$ replications, and as such, for simplicity, the results shown in all figures stem from this choice of initialization. Finally, the numerical optimization procedure for MLE sometimes reached the maximum number of iterations before converging. As in the model for positive dependence, the maximum number of iterations was set to $1000$; non-convergence within this tolerance occurred only $0.1\%$ of the time, and only for the one-step ML procedure.

\begin{figure}[H]
\centering
  \includegraphics[width=0.7\linewidth]{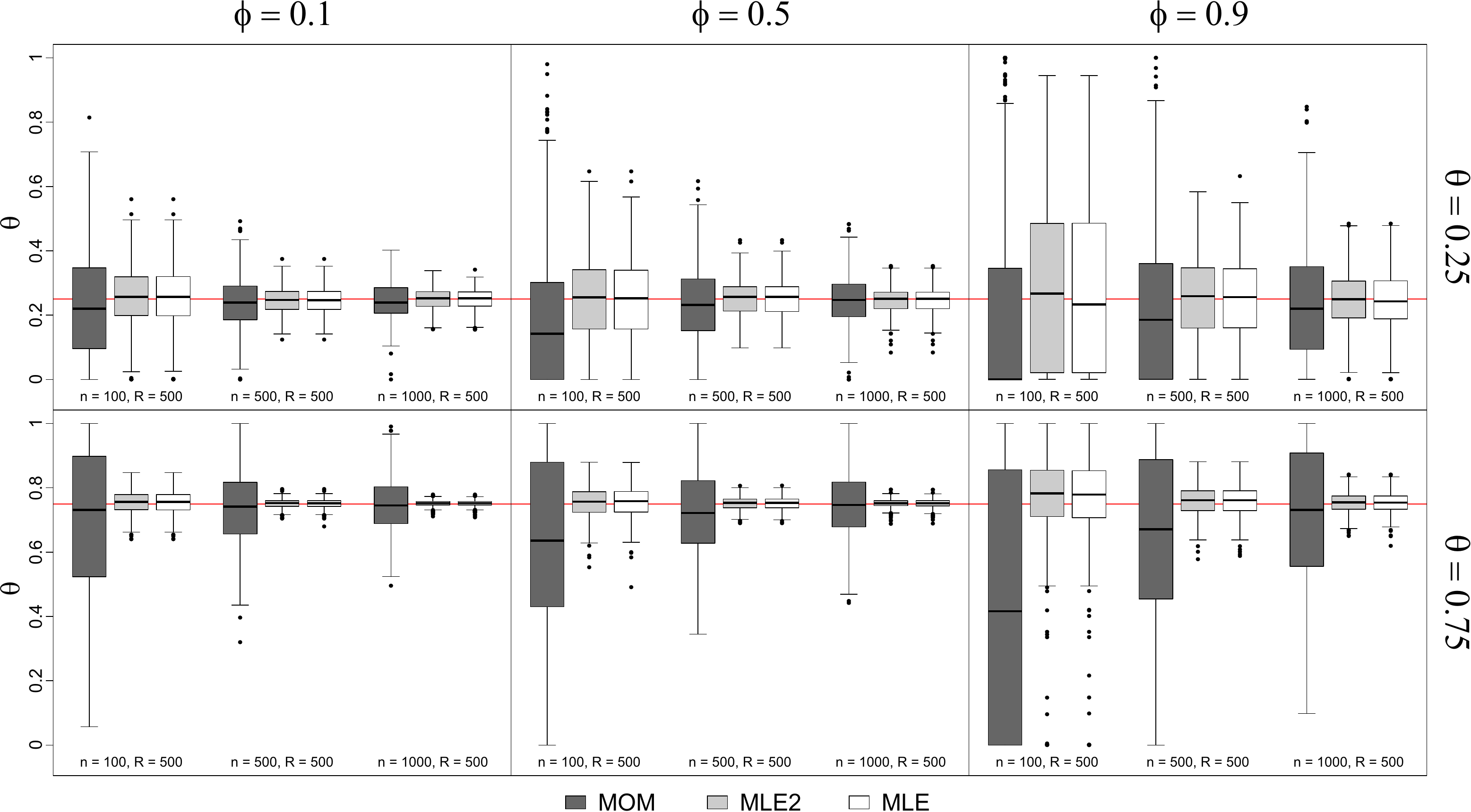}

\caption{Estimation of $\theta$ in the $\mathcal{BZIP}^-$ model across scenarios where $\Lambda=(5,10)$. The top and bottom rows respectively display the results for weaker levels of dependence (\(\theta = 0.25\)), and stronger levels of dependence $(\theta=0.75)$, while the columns display increasing zero-inflation rates (from left to right, $\phi=0.1$, $\phi=0.5$ and $\phi=0.9$). Each plot shows the results for $n \in \{100, 500, 1000\}$, across $R$ replications, with the true parameter value indicated by the red horizontal line.}
 \label{fig:neg_theta_5,10}
\end{figure}

\section{Data illustration}
\label{sec:data}

To illustrate the use of the proposed model, we explored historical data on environmental disasters in Australia. The data are taken from a public dataset (\texttt{auscathist}) in the \texttt{CASdatasets} package in $\textsf{R}$, and consists of records of catastrophic events from 1967 to 2014 in Australia. The data was reformulated in terms of the number of monthly records (for a total of $576$ months) of various types of disasters, such as cyclones, earthquakes, floods, storms, bushfires, and other types as well. In this illustration, we focus on the pair of counts stemming from the number of monthly flood and storm occurrences to assess the model's performance under positive dependence, and the pair of flood and bushfire monthly occurrences to study the performance of the negative dependence model. Table~\ref{tab:joint-occurrence} provides a brief summary of these counts. 

\begin{table}[h!]
\centering
\begin{subtable}[t]{0.45\textwidth}
\centering
\begin{tabular}{c c c c}
\toprule
Storm\textbackslash{}Flood & 0 & 1 & 2 \\ \midrule
0 & 488 & 20 & 1 \\
1 & 37  & 17 & 0 \\
2 & 4   & 3  & 5 \\
3 & 1   & 0  & 0 \\ \bottomrule
\end{tabular}
\caption{Joint monthly occurrences of storm and flood.}
\label{tab:storm-flood}
\end{subtable}
\hfill
\begin{subtable}[t]{0.45\textwidth}
\centering
\begin{tabular}{c c c c}
\toprule
Bushfire\textbackslash{}Flood & 0 & 1 & 2 \\ \midrule
0 & 508 & 39 & 6 \\
1 & 19  & 1  & 0 \\
2 & 3   & 0  & 0 \\ \bottomrule
\end{tabular}
\caption{Joint monthly occurrences of bushfire and flood.}
\label{tab:bushfire-flood}
\end{subtable}
\caption{Joint occurrences of different types of disasters monthly.}
\label{tab:joint-occurrence}
\end{table}

The $\mathcal{BZIP}^+$ model was fit to the joint occurrences of storm and flood events. The resulting MoM and MLE2 estimates are summarized in table \ref{tab:bootstrap_PD}, and provide the mean and standard deviation across $500$ bootstrap replications, along with corresponding $95\%$ boostrap confidence interval. Note that, similarly to what was observed in the simulation study in Section~\ref{sec:simulations}, there were some bootstrap replications (0.20\%) that yielded a sample with one margin consisting of only the values of $0$ or $1$; such cases were removed so that the results in fact summarize $499$ bootstrap replications. Additionally, there were some instances wherein the estimate for $\phi$ was negative (0.80\% and 0.2\% of the replications for MoM and MLE2, respectively). In both cases, the estimate of $\phi$ is set to zero for the analysis.

The $\mathcal{BZIP}^-$ was fit to the pair of bushfire and storm occurrences. Results are summarized in table \ref{tab:bootstrap_ND}, again providing information on the MoM and MLE2 estimates across 500 replications. Similarly to the case for positive dependence, some bootstrap replications led to numerical issues. Notably, 5.2\% replications resulted in at least one margin containing only 0's and 1's, so that in removing these cases, the table \ref{tab:bootstrap_ND} is in fact based on 476 bootstrap samples. Furthermore, in 6.8\% of replications, the sample covariance extended beyond that permissible in the model, such that the estimate of $\theta$ was set to $0$ (4.8\%) or $1$ (2\%), as appropriate. Note that these instances are retained in the reported results. Negative estimates of $\phi$ occur in 0.80\% of the samples using MoM and in 0.40\% using MLE2, and are similarly forced to zero for the analysis.

Both the models for positive and negative dependence suggest a significant zero-inflation, with confidence intervals excluding the value of $0$, thereby rendering the assumption of Poisson counts in the $\mathcal{BP}$ models of \cite{genest2018new} inadequate. The estimated dependence parameter, $\theta$, in both cases suggests that there is significant dependence between the counts, and as such, the model of \cite{liu2015type} would be inappropriate as well. Overall, the proposed model is able to capture both the surplus of zeros exhibited by the data, as well as the dependence between the pairs, both in the case of positive and negative associations explored here. 

\begin{table}[h!]
\centering
\begin{tabular}{lccccc}
\toprule
 & \multicolumn{2}{c}{MoM} && \multicolumn{2}{c}{MLE} \\ 
 & Estimation & 95\% CI && Estimation & 95\% CI \\ 
\cline{2-3}\cline{5-6}
$\lambda_1$ & 0.3635\,(0.0934) & (0.35, 0.38)& & 0.3248\,(0.0862) & (0.31, 0.34) \\
$\lambda_2$ & 0.2212\,(0.0782) & (0.22, 0.23)& & 0.2112\,(0.0548) & (0.21, 0.22) \\
$\theta$    & 0.4618\,(0.0990) & (0.44, 0.49)& & 0.4484\,(0.0950) & (0.42, 0.47) \\
$\phi$      & 0.5620\,(0.1340) & (0.54, 0.59)& & 0.5470\,(0.1173) & (0.52, 0.58) \\
\bottomrule
\end{tabular}
\caption{Parameter estimates for the $\mathcal{BZIP}^+$ model based on the method of moments and maximum likelihood estimation for the pair of flood and storm occurrences. The estimated values are provided, along with standard errors in parentheses, as well as a 95\% bootstrap confidence interval.}
\label{tab:bootstrap_PD}
\end{table}

\begin{table}[H]
\centering
\begin{tabular}{lccccc}
\toprule
 & \multicolumn{2}{c}{MoM} && \multicolumn{2}{c}{MLE} \\ 
 & Estimation & 95\% CI && Estimation & 95\% CI \\ 
\cline{2-3}\cline{5-6}
$\lambda_1$ & 0.2328\,(0.1003) & (0.23, 0.24)& & 0.1519\,(0.0582) & (0.15, 0.16) \\
$\lambda_2$ & 0.2218\,(0.0798) & (0.22, 0.23)& & 0.2904\,(0.0856) & (0.28, 0.30) \\
$\theta$    & 0.8921\,(0.1176) & (0.82, 0.96)& & 0.9082\,(0.1164) & (0.84, 0.98) \\
$\phi$      & 0.6452\,(0.1333) & (0.61, 0.68)& & 0.6654\,(0.1167) & (0.62, 0.71) \\
\bottomrule
\end{tabular}
\caption{Parameter estimates for the $\mathcal{BZIP}^-$ model based on the method of moments and maximum likelihood estimation for the pair of flood and bushfire occurrences. The estimated values are provided, along with standard errors in parentheses, as well as a 95\% bootstrap confidence interval.}
\label{tab:bootstrap_ND}
\end{table}

\section{Multivariate extensions}
\label{sec: multivariateEX}

While the focus of the present paper is on the bivariate case, direct extensions to higher dimensions are possible, particularly in the model for positive dependence. Indeed, consider the same construction in \eqref{equ:bzip}, but with $\mathbf{X}=(X_1,\ldots,X_d)$ and $\mathbf{T}=(T_1,\ldots,T_d)$ for arbitrary dimension $d \geq 2$. Assuming $\mathbf{T}$ follows the multivariate comonotonic shock model $\mathcal{MP}(\Lambda,\theta)$ of \cite{Schulz/Genest/Mesfioui:2021} allows to generate positively correlated zero-inflated Poisson random variables. In particular, for $\mathbf{T} \sim \mathcal{MP}(\Lambda,\theta)$, with $\Lambda=(\lambda_1,\ldots,\lambda_d) \in (0,\infty)^d$ and $\theta \in (0,1)$, one can write
\[
(T_1,\ldots,T_d) = \left(G_{(1-\theta)\lambda_1}^{-1}(V_1), \ldots, G_{(1-\theta)\lambda_d}^{-1}(V_d) \right) + \left(G_{\theta\lambda_1}^{-1}(U), \ldots, G_{\theta\lambda_d}^{-1}(U) \right),
\]
where $V_1,\ldots,V_d,U$ are independent and identical $\mathcal{U}(0,1)$ random variables. Similar expressions for the corresponding joint PMF and CDF follow from the bivariate setting, with
$$
h^+_{\Lambda,\phi,\theta}(x_1,\ldots,x_d) = \left\{\phi+(1-\phi) f^+_{\Lambda,\theta}(0,\ldots,0)\right\}^{\delta_1 \times \cdots \times \delta_d} \times \left\{(1-\phi)f^+_{\Lambda,\theta}(x_1,\ldots,x_d)\right\}^{1-\delta_1 \times \cdots \times \delta_d}
$$
and $$
H^+_{\Lambda,\phi,\theta}(x_1,\ldots,x_d) = \phi \Delta_1 \cdots \Delta_d + (1-\phi) F^+_{\Lambda,\theta}(x_1,\ldots,x_d),
$$
where $\delta_j = \mathbf{1}(x_j=0)$ and $\Delta_j = \mathbf{1}(x_j \geq 0)$, for $j=1,\ldots,d$, and $f^+_{\Lambda,\theta}$ and $F^+_{\Lambda,\theta}$ respectively denoting the probability mass function and distribution function for the latent counts $\mathbf{T} \sim \mathcal{MP}(\Lambda,\theta)$.

Adapting the estimation procedures to higher dimensions is rather straightforward. Likelihood based estimation, including the EM algorithm with the IFM simplification, can be generalized to dimension $d>2$ as the d-variate version of $f^+_{\Lambda,\theta}$ closely resembles that of the bivariate setting, full details are given in \cite{Schulz/Genest/Mesfioui:2021}. As previously noted, in high dimensional settings, the EM algorithm (with IFM implementation) will likely be convenient for finding the maximum likelihood estimators as the parameters $\phi$ and $\lambda_1,\ldots,\lambda_d$ can be iteratively updated according to the closed-form expressions, as provided in Section~\ref{sec:EM}, while updating $\theta$ then only involves a univariate optimization. Some adaptations of the method of moments estimators are required. Namely, the estimator $\check{\phi}$ will now be determined in terms of the average of $d$ estimates, that is, $\sum_{j=1}^d \check{\phi}_j/d$. Analogously to what is done in \cite{Schulz/Genest/Mesfioui:2021}, estimation of $\theta$ can be based on matching all pairwise sample moments, leading to a set of ${d \choose 2}$ estimates $\check{\theta}_{jk}$, which can then be averaged. 

Note that extending the model for negative dependence to higher dimensions is not as straightforward. As commented by \cite{Schulz/Genest/Mesfioui:2021}, the lower Fréchet--Hoeffding bound (i.e., the case of counter-monotonicity) is not, in general, a proper distribution function in dimensions greater than 2, except in certain specific settings. As such, defining a multivariate model based on a d-variate counter-monotonic shock vector is not obvious. The discussion of the multivariate model is left brief, and a full exploration of the case where $d>2$ is left for future work. 

\section{Conclusion}
\label{sec:conclusion}

This work presents a new family of bivariate zero-inflated Poisson models allowing for both positive and negative associations. Several distributional properties are derived, notably the covariance structure, which is shown to allow for considerable flexibility in capturing varying degrees of dependence. Various estimation approaches are detailed and further explored through extensive simulations. Finally, the models for both positive and negative dependence are shown to be useful in modeling real count data exhibiting excess zeros. 

While not explicitly addressed in the present work, extending the $\mathcal{BZIP}$ models to incorporate covariate effects would be straightforward. Indeed, a regression-based version of the model could be defined by reformulating the parameters in terms of a set of covariates $\mathbf{C}$. For example, one could consider setting 
\begin{align*}
    \lambda_j &= \exp(\mathbf{C} \boldsymbol{\beta}), \quad j\in \{1,2\} \\
    \theta &= \exp(\mathbf{C} \boldsymbol{\alpha})/\left\{ 1 + \exp(\mathbf{C} \boldsymbol{\alpha})\right\} \\
    \phi &= \exp(\mathbf{C} \boldsymbol{\gamma})/\left\{ 1 + \exp(\mathbf{C} \boldsymbol{\gamma})\right\}.
\end{align*}
The regression coefficients could then be estimated using maximum likelihood estimation.

By construction, the proposed formulation relies on a common latent Bernoulli variable which acts on both margins. While there are certainly many contexts wherein a common inflation factor is intuitive, this may sometimes be too restrictive in modeling data which exhibit component-specific inflation rates. Extending the model to allow for margin-specific zero-inflation would allow for further flexibility in the model, building on the work of, e.g., \cite{li1999multivariate} and \cite{wu2023multivariate}. 
We leave such explorations for future work.

\appendix
\section{Appendix}

\subsection{Asymptotic properties of method of moments estimators}\label{app:MM}

Recall from Section~\ref{sec:MM} that the method of moments estimation based on the marginal specifications leads to 
\[
\check{\lambda}_j=\frac{\sum_{i=1}^n X_{ij}^2}{n\bar{X}_j} -1, \quad \check{\phi}_j = 1-\frac{\bar{X}_j}{\check{\lambda}_j}, \quad j=1,2.
\]
Standard statistical results ensure that both the above estimators are consistent and asymptotically normal. For $(X_{i1},X_{i2}) \overset{iid}{\sim} \mathcal{BZIP}(\Lambda,\theta,\phi)$, for $i=1,\ldots,n$, the central limit theorem then implies that as $n \to \infty$
\[
\frac{\frac{1}{n} \sum_{i=1}^n X_{ij}^2 - (1-\phi)\lambda_j(1+\lambda_j)}{\sigma/\sqrt{n}} \rightsquigarrow \mathcal{N}(0,1), 
\]
where $\sigma^2 = \var(X_j^2)=(1-\phi)\{ \phi \lambda_j^4 +2(2+\phi)\lambda_j^3+(8+\phi)\lambda_j^2 +\lambda_j \}$. Since $\bar{X}_j \overset{a.s.}{\to} (1-\phi)\lambda_j$, by Slutsky's theorem one can establish that as $n \to \infty$
\[
\check{\lambda}_j \rightsquigarrow \mathcal{N}(\lambda_j, \xi^2/n)
\]
where $\xi_j = \sigma / \{(1+\phi)\lambda_j\}$, and accordingly, that $\check{\lambda}_j \overset{p}{\longrightarrow} \lambda_j$.

Using similar arguments, it can be shown that $\sqrt{n}(\check{\phi}_j-\phi)$ is asymptotically normal with mean $0$ and variance $(1-\phi)(1+\phi\lambda_j)/\lambda_j$. The asymptotic properties of the MoM estimator, $\check{\phi}=(\check{\phi}_1+\check{\phi_2})/2$, then follows, for example, by the Cramér-Wold theorem, which ensures that $\check{\phi}$ is also asymptotically normal with 
\[
\sqrt{n}(\check{\phi}-\phi) \rightsquigarrow \mathcal{N}(0,\zeta^2)
\]
where 
\[
\zeta^2 = \frac{(1-\phi)}{4}\left\{ \frac{(1+\phi\lambda_1)}{\lambda_1}+ \frac{(1+\phi\lambda_2)}{\lambda_2} \right\} + \frac{n}{2}\cov(\check{\phi}_1,\check{\phi}_2). 
\]
Using the delta method allows for further simplifications. Indeed, let $\mathbf{M} = (X_{1},X_{1}^2,X_{2},X_{2}^2)$, and let $\mathbf{M}_n$ denote the corresponding vector of first and second sample moments based on a random sample $(X_{11},X_{12}),\ldots,(X_{n1},X_{n2})$, i.e., 
\[
\mathbf{M}_n= \left( \frac{1}{n}\sum_{i=1}^nX_{i1},\frac{1}{n}\sum_{i=1}^nX_{i1}^2,\frac{1}{n}\sum_{i=1}^nX_{i2},\frac{1}{n}\sum_{i=1}^nX_{i2}^2 \right)
\]
By the multivariate central limit theorem, 
\begin{equation*}
    \sqrt{n}(\mathbf{M}_n - \Omega) \rightsquigarrow \mathcal{N}(\mathbf{0},\Sigma_M)
\end{equation*}
where $\Omega$ denotes the mean vector $\E(\mathbf{M})$,  given by
\[
\Omega=\{(1-\phi)\lambda_1 , (1-\phi)\lambda_1(1+\lambda_1), (1-\phi)\lambda_2, (1-\phi)\lambda_2(1+\lambda_2)\}
\]
and the asymptotic covariance matrix $\Sigma_M$ is given by 
\begin{equation*}
\cov(\mathbf{M}) = 
\begin{bmatrix}
\var(X_{1}) & \cov(X_{1}, X_{1}^2) & \cov(X_{1}, X_{2}) & \cov(X_{1}, X_{2}^2) \\
\cov(X_{1}^2, X_{1}) & \var(X_{1}^2) & \cov(X_{1}^2, X_{2}) & \cov(X_{1}^2, X_{2}^2) \\
\cov(X_{2}, X_{1}) & \cov(X_{2}, X_{1}^2) & \var(X_{2}) & \cov(X_{2}, X_{2}^2) \\
\cov(X_{2}^2, X_{1}) & \cov(X_{2}^2, X_{1}^2) & \cov(X_{2}^2, X_{2}) & \var(X_{2}^2)
\end{bmatrix} .
\end{equation*}
The above can be further simplified. In particular, for $j=1,2$, one has that 
\begin{align*}
    \var(X_j) &= (1-\phi)\lambda_j(1+\phi\lambda_j)\\
    \var(X_j^2) &= (1-\phi)\lambda_j\{\phi\lambda_j^3 + 2(2+\phi)\lambda_j^2 + (6+\phi)\lambda_j + 1\} \\
    \cov(X_j,X_j^2) &= (1-\phi) \lambda_j \{\phi\lambda_j^2 + (\phi+2)\lambda_j + 1\} 
\end{align*}
Furthermore, in the $\mathcal{BZIP}$ model 
\begin{align*}
    \cov(X_1,X_2) &= (1-\phi) \cov(T_1,T_2) + \phi(1-\phi)\lambda_1\lambda_2\\
    \cov(X_1^2,X_2) &= (1-\phi) \cov(T_1^2, T_2) + \phi(1-\phi)\lambda_1\lambda_2(1+\lambda_1)\\
    \cov(X_1,X_2^2) &= (1-\phi) \cov(T_1, T_2^2) + \phi(1-\phi)\lambda_1\lambda_2(1+\lambda_2)\\
    \cov(X_1^2,X_2^2) &= (1-\phi) \cov(T_1^2, T_2^2) + \phi(1-\phi)\lambda_1\lambda_2(1+\lambda_1)(1+\lambda_2)
\end{align*}
where expressions for the covariances related to the Poisson pairs $(T_1,T_2)$ can be derived from the probability generating function, as given in \cite{genest2018new}. 

Let $M_{1j} = \bar{X}_j$, and $M_{2j}=\frac{1}{n} \sum_{i=1}^n X_{ij}^2$, so that $\mathbf{M}_n = (M_{11}, M_{21},M_{12}, M_{22})$. The MoM estimator of $\phi$ is a function of the components of $\mathbf{M}_n$, viz.
 \begin{equation*}
     \check{\phi} = g(\mathbf{M}_n) = g(M_{11}, M_{21},M_{12}, M_{22}) = \frac{1}{2} \left(2-\frac{M_{11}^2}{M_{21} - M_{11}} -\frac{M_{12}^2}{M_{22} - M_{12}} \right),
 \end{equation*}
and $\phi = g(\Omega)$. By the delta method, one has that  
\[
    \sqrt{n}(\check{\phi} - \phi) \rightsquigarrow \mathcal{N}(0, \nabla g(\Omega)^\top \Sigma_M \nabla g(\Omega))
\]
where $\nabla g(\Omega)$ denotes the gradient of the function $g$, evaluated at $\Omega$. In particular, we have that the partial derivatives of $g(m_{11},m_{21},m_{12},m_{22})$ are given by
\begin{align*}
    \frac{\partial \, g}{\partial \,m_{1j}} = \frac{1}{2} \left\{ \frac{m_{1j}(m_{1j}-2m_{2j})}{(m_{2j}-m_{1j})^2}\right\}, \quad \frac{\partial \, g}{\partial \, m_{2j}} = \frac{1}{2} \left( \frac{m_{1j}}{m_{2j}-m_{1j}} \right)^2, \text{ for } j=1,2,
\end{align*}
thereby yielding
\[
\nabla g(\Omega)^{\top} = \left\{ \frac{-(1+2\lambda_1)}{2\lambda_1^2} , \frac{1}{2 \lambda_1^2} , \frac{-(1+2\lambda_2)}{2\lambda_2^2} , \frac{1}{2 \lambda_2^2}\right\}
\]

In general, the asymptotic variance of the MoM estimator $\check{\phi}$, i.e., $\nabla g(\Omega)^\top \Sigma_M \nabla g(\Omega)$, has a complex form, which can be estimated via bootstrap. However, simplifications are possible in certain settings. For example, when $\theta=0$, it follows that
\begin{align*}
    \cov(X_1,X_2) &= \phi(1-\phi)\lambda_1\lambda_2 ,\\
    \cov(X_1^2,X_2) &=  \phi(1-\phi)\lambda_1\lambda_2(1+\lambda_1) ,\\
    \cov(X_1,X_2^2) &=  \phi(1-\phi)\lambda_1\lambda_2(1+\lambda_2) , \\
    \cov(X_1^2,X_2^2) &= \phi(1-\phi)\lambda_1\lambda_2(1+\lambda_1)(1+\lambda_2).
\end{align*}
In this case, it can be shown that the asymptotic variance simplifies to $(1-\phi) ( \lambda_1^{-2}/2 + \lambda_2^{-2}/2 + \phi )$.

The MoM estimator $\check{\theta}$ is the unique solution to $s_{\check{\Lambda},\check{\phi}}(\theta)=S_{12}$, provided that $S_{12}$ falls within the permissible ranged as determined by the bounds $s_{\check{\Lambda},\check{\phi}}(0)$ and $s_{\check{\Lambda},\check{\phi}}(1)$. The asymptotic properties of $\check{\theta}$ can be established using similar arguments to \cite{genest2018new}. First, as noted in, e.g., Theorem~8 of \cite{Ferguson:1996}, the sample covariance is asymptotically normal, with $\sqrt{n}\left\{S_{12}-s_{\Lambda,\theta}(\theta)\right\} \rightsquigarrow \mathcal{N} (0,\tau^2)$ where $\tau^2=\var\left[\{X_1-(1-\phi)\lambda_1\}\{X_2-(1-\phi)\lambda_2\} \right]$. From this, applying the delta method allows to determine the asymptotic behavior of $\check{\theta}$. Specifically, as $n\rightarrow \infty$, 
\[
\sqrt{n}(\check{\theta}-\theta) \rightsquigarrow \mathcal{N}(0,\left\{ \gamma^{\prime}(\sigma_{12})\right\}^2 \tau^2)
\]
where $\sigma_{12}=\cov(X_1,X_2)$, $\gamma$ is the inverse covariance function, i.e.$\gamma : \theta \mapsto s^{-1}_{\Lambda,\phi}(\theta)$, and $\gamma^{\prime}$ is the corresponding derivative. Note that in certain settings, further simplifications may be possible. For example, when $\phi=0$, the $\mathcal{BZIP}$ model simplifies to the bivariate Poisson model of \cite{genest2018new}. In the latter, the authors provide some simplified expressions, particularly in the case where $\lambda_1=\lambda_2$.

\section*{Funding}

This work was supported by the Natural Sciences and Engineering Research Council of Canada (JS: RGPIN-2022-05118, JFP: DDG-2023-00030) and IVADO (JFP: PRF-2019-3055954398).

\bibliographystyle{apalike}

\bibliography{mybibfile}

\end{document}